\documentclass[%
superscriptaddress,
 amsmath,amssymb,
 aps,
 prl,
 reprint,
floatfix,
]{revtex4-1}
\usepackage{graphicx,xcolor}
\definecolor{darkblue}{RGB}{0,0,150}
\definecolor{nightblue}{RGB}{0,0,100}

\usepackage{mathrsfs,dsfont,mathtools}

\usepackage{ulem}

\usepackage{dcolumn}
\usepackage{bm}
\usepackage[
colorlinks,
citecolor=darkblue,
linkcolor=darkblue,
urlcolor=nightblue]{hyperref}

\usepackage[english]{babel}
\usepackage[babel,kerning=true,spacing=true]{microtype}

\usepackage{feynmp-auto}
\newcommand{\resp}[1]{\textcolor{black}{#1}}

\AtBeginDocument{}
\begin{document}

\title{Tunable Spatiotemporal Orders in Driven Insulators
}

\author{Daniel Kaplan}
\email{d.kaplan1@rutgers.edu}
\address{Center for Materials Theory, Department of Physics and Astronomy,
Rutgers University, Piscataway, NJ 08854, USA}%
\author{Pavel A. Volkov}
\email{pavel.volkov@uconn.edu}
\address{Department of Physics, University of Connecticut, Storrs, CT 06269, USA}
\author{Ahana Chakraborty}
\address{Center for Materials Theory, Department of Physics and Astronomy,
Rutgers University, Piscataway, NJ 08854, USA}
\affiliation{Department of Physics and Astronomy, Louisiana State University, Baton Rouge, Louisiana 70803, USA}
\author{Zekun Zhuang}
\address{Department of Physics, University of Wisconsin-Madison, Madison, Wisconsin 53706, USA}
\author{Premala Chandra}
\email{premala@physics.rutgers.edu}
\address{Center for Materials Theory, Department of Physics and Astronomy,
Rutgers University, Piscataway, NJ 08854, USA}
\date{\today}
\begin{abstract}
We show that driving optical phonons above a threshold fluence induces spatiotemporal orders, where material properties oscillate at an incommensurate wavevector $q_0$ in space and at half the drive frequency in time. The order is robust against temperature on timescales much larger than the lifetime of the excited modes and can be accompanied by a static $2q_0$ modulation. We make predictions for time-resolved diffraction and provide estimates for candidate materials. Our results show the possibility of using THz waves in solids to realize tunable incommensurate order on the nanoscale.
\end{abstract}

\maketitle

\paragraph{Introduction---.} Driven condensed matter phases are attracting tremendous interest \cite{basov2017,de2021colloquium,Bloch22}
 due to the rapid development of ultrafast optics 
\cite{hebling2008,emma2010,basov2017,de2021colloquium,Bloch22,Buzzi18,Disa21,Bloch22,Henstridge22} and proposals of exotic non-eqilibrium phases, such as time crystals \cite{Khemani19,Zaletel23,Shapere12,Yao20,Wilczek12,Khemani16}. The
use of terahertz pulses has led to the observation of several light-induced transitions in different quantum materials \cite{basov2017,nova2019,Disa21,Henstridge22,li2019terahertz,disa2023photo,cheng2023}.
Recent advances in time-resolved diffraction probe spatial features of transient orders on time-scales set by the width of a pumping pulse ($10^{-12} ~\textrm{sec}$) \cite{Fechner2024,Orenstein24}. 

The conventional mechanism for such light-driven 
order involves the dynamical softening of a near-critical mode, resulting in a symmetry-breaking transition \cite{subedi2014,Subedi2015,Subedi2017,Zhuang23,Puviani2018,Radaelli18,rubio2022, Pueyo22, Pueyo23, Kuzmanovskii24, Khalsa24}; this approach is appropriate for materials close to instabilities where accessible fluences are sufficient to affect the ground state. 
As the nature of the transient order is set by the system's pre-existing soft modes, its characteristic features
are generally not tunable \cite{nova2019,li2019terahertz,disa2023photo} and the transient order is commensurate with the underlying lattice.

\begin{figure}[ht]
\centering
 \includegraphics[width=1.0\columnwidth,page=2]{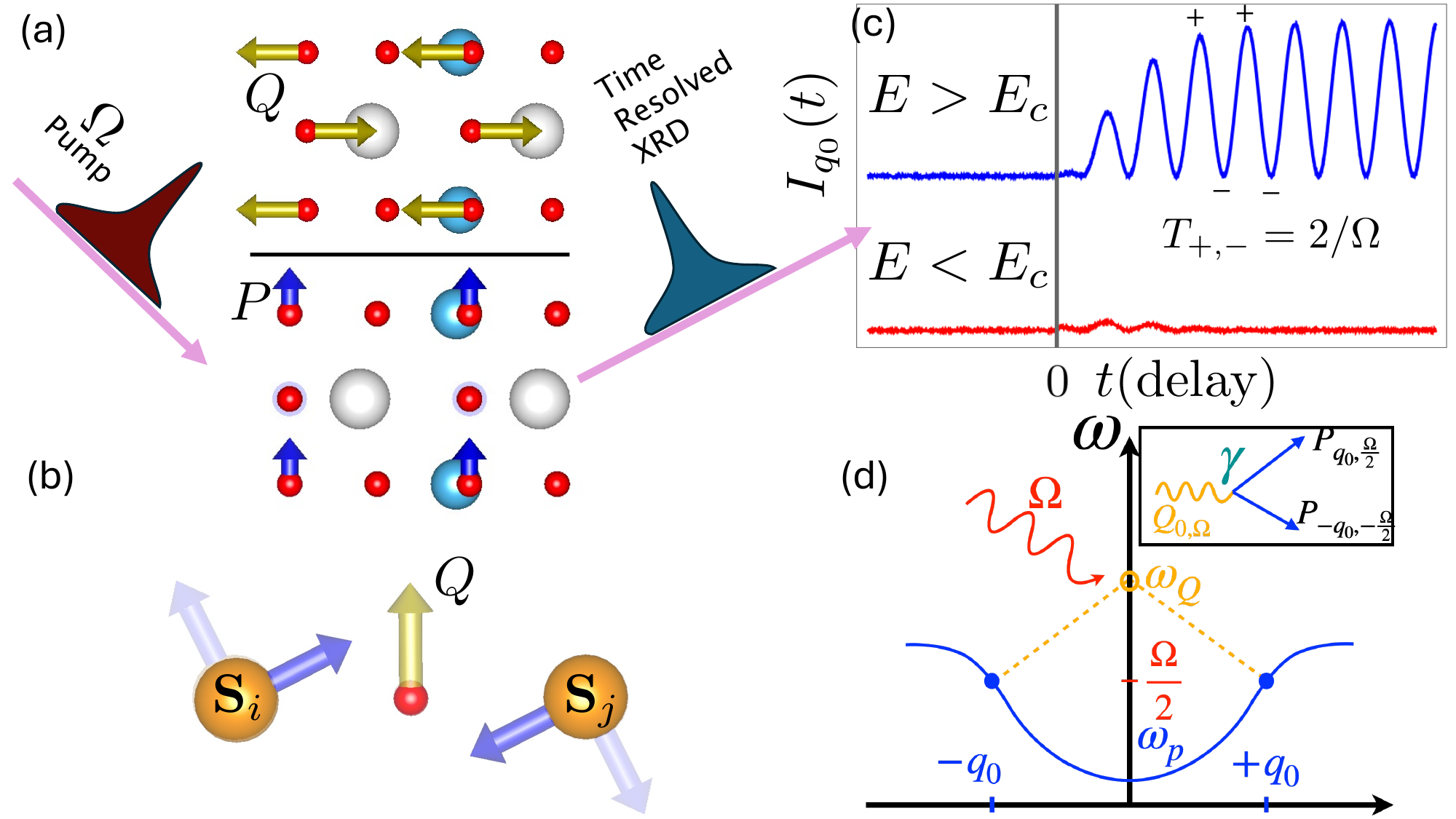}
    \caption{Schematics and experimental signatures of the induced spatio-temporal order. (a) In a pump-probe setup, an optical phonon $Q$ is resonantly driven at frequency $\Omega$ and is nonlinearly coupled to a low energy mode $P$. Parametric resonance of $P$ at a finite wavevector (see (d)) drives the formation of spatiotemporal order probed with X-Ray diffraction (XRD).
    (b) Similar to (a), in a magnetic system with strong spin-orbit coupling (SOC), $Q$ generates spin
   order (here $P$ is the magnetization) via an induced dynamic Dzyaloshinskii-Moriya (DM) interaction. (c) In time-resolved XRD, the intensity of the induced order at wavevector $q_0$ will vary in time as a function of the delay from the pump. The amplitude of the mode $P_{q_0}$ alternates between maxima and minima with a period $T_{+,-} = (\Omega/2)^{-1}$. The mode condenses only if the drive amplitude $E$, $E>E_c$ (blue curve) and decays exponentially for $E<E_c$ (red curve). (d) Sketch of the parametric resonance mechanism. $Q$ is coupled to $P$ via a cubic interaction with strength $\gamma$. The parametric resonance condition selects the momentum $q_0$ for which the dispersion of $P$, $
    \omega_P(q_0) = \Omega/2$ (see main text). 
    }
     \vspace{-1.8em}
    \label{fig:schematic}
\end{figure}

In this work, we present an alternative pathway to light-induced phase transitions that results in tunable spatiotemporal orderings due to the interplay of nonlinearity, parametric resonance and spatial dispersion.
We consider a polar phonon $Q$ with natural frequency $\omega_Q$, resonantly driven at $\Omega \approx \omega_Q$; $Q$ is nonlinearly coupled to a low-energy excitation $P$, such as a phonon or magnon (Fig. \ref{fig:schematic} (a,c,d)). For fluences above a threshold set by the linewidths of $Q$ and $P$, a steady state develops, where the $P$ degree of freedom 
oscillates in time with frequency $\Omega/2$ and in space with wavevector $q_0$ determined by its dispersion (Fig. \ref{fig:schematic} (c)). 
Our numerical results indicate that the resulting spatiotemporal order is generally stable against temperature, and we predict its signatures in time-resolved diffraction experiments on candidate materials. 
\allowdisplaybreaks
\paragraph{Model and Equations of Motion---.} The key features for inducing spatiotemporal order are shown
in Fig.~\ref{fig:schematic}(c): optical ($Q_i(r,t)$) and $P_i(r,t)$ (of any symmetry) modes
with frequencies $\omega_Q(q)$ and $\omega_P(q)$ respectively such that $\omega_P(0) < \omega_Q(0)/2$ .
Parametric resonance is driven by a cubic coupling as shown in Fig.~\ref{fig:schematic}(d). Above a critical fluence, an additional Bragg peak at $q_0$ is expected in time-resolved diffraction experiments \cite{Orenstein24,de2023ultrafast,huang2024hard,Fauque22,johnson2023ultrafast} whose amplitude varies at frequency $\frac{\Omega}{2}$. $q_0$ is set by the dispersion of $P$, with the condition $\omega_P(q_0) = \Omega/2$.
\begin{figure*}
    \centering
    \includegraphics[width=\textwidth]{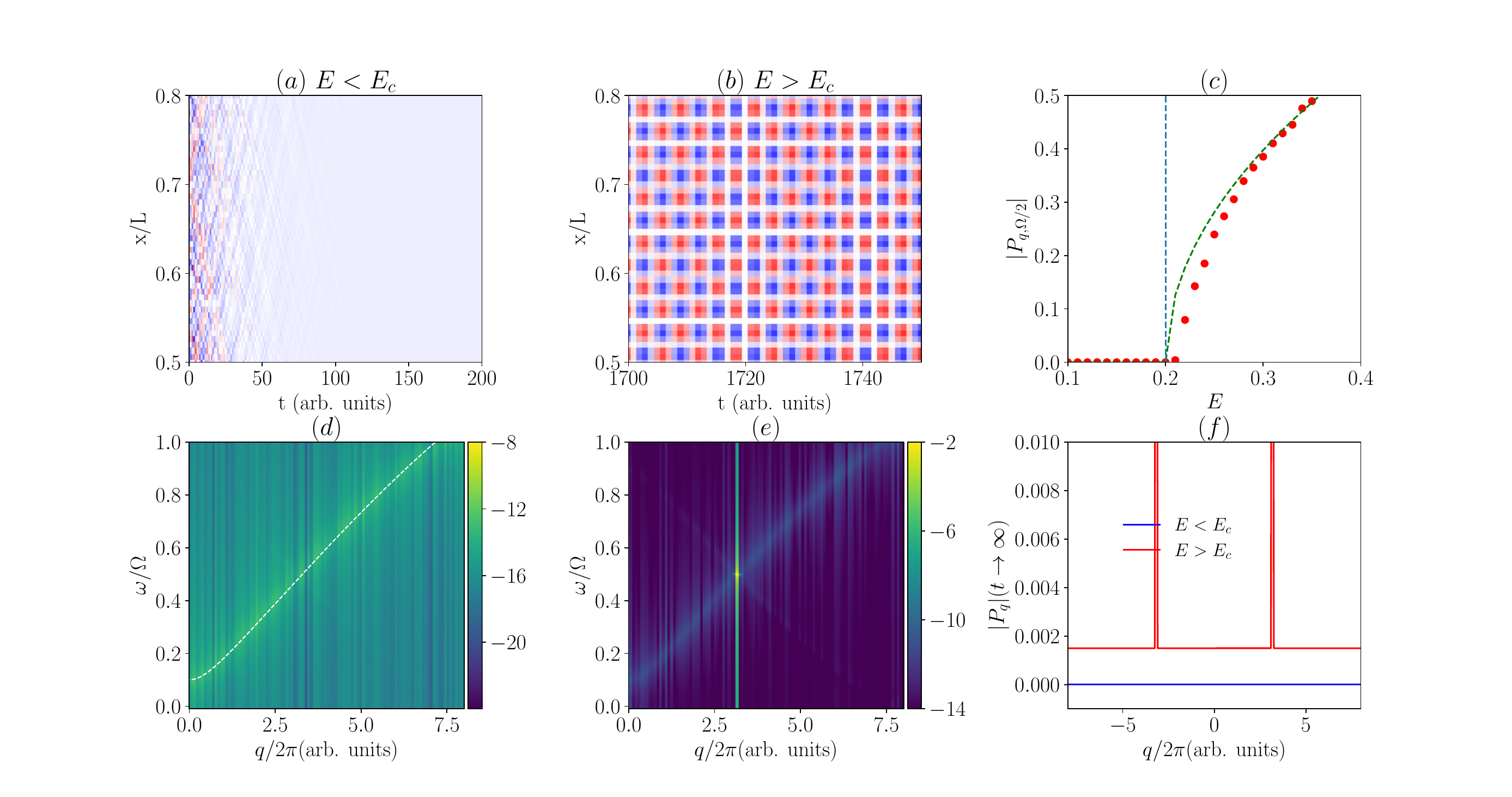}
    \caption{General solution of the coupled nonlinear equations at $T=0$. (a) Spatial distribution of the mode $P(x,t)$ for electric field strength $E$ below the threshold $E_c$. (b) Symmetry-broken phase shown for $E > E_c$,  zooming in on $x/L$ and $t \gg 1/\beta$ (within the steady state); $P$ oscillates in both space and time signifying spatio-temporal order. This order is stable on all scales for times greater than $1/\beta$. (c) Amplitude of the parametrically excited mode $P_{q, \Omega/2}$ as a function of $E_c$. For the parameters selected here, $P_{q,\Omega/2}$ appears at $E\approx 0.2$. The blue dashed line is the critical value calculated from Eq.~\eqref{eq:critical_value}. The green dashed line is the amplitude of the condensed mode from Eq.~\eqref{eq:critical_value}. The small deviation from the analytical result in numerics is due to the finite integration time in the simulation. The vanishing of $P_{q,\Omega/2}$ is continuous. (d) Log of the spectral function $\textrm{ln} |S(q,\omega)|$ (see main text). For clarity, the dispersion of $P(x,t)$ is superimposed with a dashed white line. For $E< E_c$, the amplitude is exponentially suppressed, but signatures of the bare band structure of $P(x,t)$ are visible. (e) $\textrm{ln} |S(q,\omega)|$ for $E>E_c$, showing the sharp peak at $q \approx 3.1$ and $\omega = \Omega/2$ confirming the parametric resonance. (f) Fourier transform of $P(x,t)$ in $q$ at late times. Clearly visible $q \approx 3.1$ peak appears only for $E>E_c$. Parameters used here are: $\omega_P = 0.2$, $\Omega=\omega_Q = 2.0$, $\beta = 0.1$, $v_P = 4\cdot10^{-3}, \gamma = 0.1, \alpha_{Q,P}'=0.0001$.}
    \label{fig:fig1}
     \vspace{-2.0em}
\end{figure*}
The cubic coupling we described above can generically take the following form \cite{Subedi2015,subedi2021light}, 
\begin{align}
 V_{\textrm{c}}(Q,P) =  \xi_{ijk} Q_i P_j P_k + \xi'_{ijk} Q_{i}Q_j P_k + \delta_{ijkl} (Q_i P_j)\partial_{k} P_l.
 \label{eq:cubic_pot}
\end{align}
We also allow for nonlinear self-interaction in the modes,
\begin{align}
      V_{nl}(Q,P) = \alpha_Q Q^3 + \alpha_Q' Q^4+\alpha_P P^3 + \alpha_P' P^4.
     \label{eq:nonlin}
\end{align}
We set $\xi'_{ijk} = 0$ and discuss the coupling to the electric field and lower order terms in the Supplementary Material (SM). As $Q_{i}Q_{j} P_{k}, (Q_i P_j)\partial_{k} P_l$ have the form $g_{ijk} (q) Q_{i} Q_{j} P_{k}$ in momentum space, they lead to qualitatively similar physics. The first ($g(q) = \textrm{const.}$) is allowed only for polar systems, while the latter is always present, and is relevant particularly for magnetic systems (see Discussion and Fig.~\ref{fig:schematic}(b)). 
We begin with a concrete example of two coupled polar phonons in a ferroelectric (Fig. \ref{fig:schematic} (a)). 
For a symmetry group where $\xi_{ijk} = \gamma$, the classical equations of motion are,
\begin{align}
\notag   &\ddot{Q} - v_Q^2 \nabla^2 Q + \omega_Q^2Q + \beta \dot{Q} = \\ & \qquad \qquad E(t) -\gamma P^2-\frac{\partial V_{\textrm{nl}}(Q)}{\partial Q} + \eta(r,t), \label{eq:qmode} \\ \notag 
    &\ddot{P} - v_P^2 \nabla^2 P + \omega_P^2 P + \beta \dot{P} = \\ & \qquad \qquad  -2 \gamma P Q- 
    \frac{\partial V_{\textrm{nl}}(P)}{\partial P} +  \eta(r,t).\label{eq:pmode}
\end{align}
where dissipation and temperature are modeled with isotropic damping 
($\beta \dot{Q}, \beta \dot{P}$) and Langevin, $\eta(r,t)$ terms respectively. \resp{The Langevin term satisfies $\langle \eta(r,t)\eta(r',t') \rangle = 2\beta T \delta(t-t')\delta(r-r')$, where $T$ is the temperature; discussion of dimensions in the EOMs is left to the SM.} \resp{Here we focus on insulators with a large bandgap $\Delta \gtrsim \textrm{1eV}$ while the driving frequency is $\Omega \lesssim 100 \textrm{meV}$
with these scales directly connected with the properties of e.g. PbTiO\textsubscript{3} ($\Delta \approx \textrm{1eV}$,  $\Omega \approx 40 \textrm{meV}$)\cite{Choudhury2008}, assuming temperatures to be well below $\Delta/k_B$ supressing thermal excitations.}

\paragraph{Numerical Solutions at $T=0$---.} We solve \eqref{eq:qmode}-\eqref{eq:pmode} numerically, drawing $P(x,0), Q(x,0)$ from a uniform distribution with a small amplitude $|P(x_i,0)| < \frac{L}{N_x}$, where $N_x$ is the discretization along the $x$ axis and 
$\dot{Q}(x,0)= \dot{P}(x,0)=0$ such that 
nonlinearities are initially negligible. \resp{For simplicity, the equations are solved in the 1D case, for a field supported on $x,t$. The generalization to higher dimesions is straightforward and discussed in the sections below.}
Eqs. \eqref{eq:qmode}-\eqref{eq:pmode}
are solved using standard methods for Langevin dynamics \cite{leimkuhler2013robust}, satisfying the
CFL criterion \cite{lax2013stability} for all parameters thus guaranteeing stability of the numerical integration. \resp{Details about the numerical simulation are relegated to the Supplementary information}.
For fields $E < E_{c}$, we find that for any set of initial conditions the $P$ mode decays exponentially on a scale of $1/\beta$; a representative solution is presented in Fig.~\ref{fig:fig1}(a). 
Nonetheless, the observed dynamics contains information about the coupling between the modes and the dispersion. In particular, the structure factor $S(q,\omega) = \int \textrm{d}x \textrm{d}t P(x,t) e^{iqx} e^{i\omega t}$;  (Fig.~\ref{fig:fig1}(d)) shows a pronounced branch in frequency-momentum space reflecting the dispersion of the $P$ mode.
For $E>E_c$, a regular pattern emerges at late times (Fig.~\ref{fig:fig1}(b)), oscillating in time with frequency $\Omega/2$. This $(q_0,\frac{\Omega}{2})$ order also has a fingerprint in $S(q,\omega)$ (Fig.~\ref{fig:fig1}(e)), where an extremely strong peak at $q_0 \approx 3.1 /(2 \pi a)$ appears.
A partial structure factor analysis, $S(q,t \to \infty)$, for  $E < E_c$ and $E > E_c$ is plotted in Fig.~\ref{fig:fig1}(f), 
indicating a delta-like peak in momentum for $E>E_c$ at $q = |q_0|$. The initial-condition averaged value of $P_{q_0,\Omega/2}$ as a function of field amplitude is
presented in Fig.~\ref{fig:fig1}(c). The onset of a finite $|P_{q_0,\Omega/2}|$ above $E_c$ is continuous and follows $\sim \sqrt{E-E_c}$.

\begin{figure}
    \centering
    \includegraphics[width=\columnwidth]{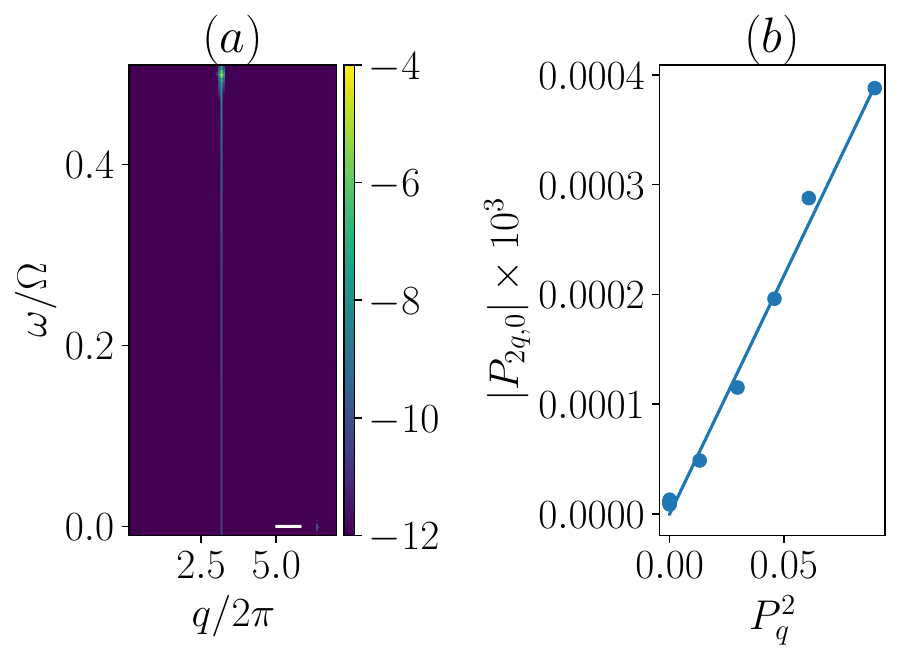}
    \caption{Finite $2q, \omega = 0$ modulation with additional nonlinearity at $T=0$. (a) Spectral function focusing on the segment with $q > 3$ and $\omega < 1/2 \Omega$. Note the additional peak at $2q \approx 6.2 / (2\pi a)$, see peak next to the white bar. The peak at $q \approx 3.1 / (2\pi a) $ is unchanged. This mode is static in time. (b) Dependence of the amplitude of the $P_{2q, \omega=0}$ mode as a function of $P_{q}$. Note that $P_{2q} \propto |P_{q}|^2$, which is plotted in Fig.~\ref{fig:fig1}(c). Solid line is the analyical solution presented in the main text. Parameters identical with Fig.~\ref{fig:fig1}(c), but with $\alpha_{P} = 0.01$. }
    \label{fig:fig3}
    \vspace{-2em}
\end{figure}

Next we introduce a non-negligible cubic nonlinearity for the $P$ mode in $V_{nl}$ and plot the resulting spectral function in Fig.~\ref{fig:fig3}(a); comparison with Fig.~\ref{fig:fig1}(e) indicates a new static ($\omega = 0$) feature at $2q_0$ whose
amplitude is weaker than that of $|P_{q,\Omega/2}|$. In Fig.~\ref{fig:fig3}(b) the scaling of this mode $|P_{2q_0,0}|$ 
as a function of fluence is displayed; $|P_{2q_0,0}| \sim (E-E_c)$, indicating a continuous transition at the same $E_c$  as $|P_{q_0,\frac{\Omega}{2}}|$ but with a higher exponent.

\paragraph{Nonlinear Parametric Resonance---.}
Inspired by the numerical results, we show that the tunable spatiotemporal order results from parametric resonance \cite{Landau76} at finite momentum. We approximate the the fields $Q,P$ by the contributions of three modes that dominate to the dynamics at $E>E_c$: $Q(x,t) \propto Q_{0,\Omega} e^{i\Omega t}+c.c.$ and $P(x,t) \approx P_{q_0, \Omega/2} e^{i\left(\frac{\Omega}{2}t-q_0 x\right)}+P_{q_0,-\Omega/2} e^{-i\left(\frac{\Omega}{2} t+q_0 x\right)} +c.c. $. 
Using this ansatz the EOMs are:
\begin{align}
       &\notag \Omega^2 Q_{0,\Omega} - \omega_Q^2 Q_{0,\Omega} -i \Omega \beta Q_{0,\Omega} = \\ & \qquad \qquad -E_0/2 +2\gamma P_{q_0,\Omega/2} P^*_{q_0,-\Omega/2}\label{eq:coupled_minimal_q} \\ \notag
       &(\Omega/2)^2 P_{q_0,\Omega/2} - (\omega_P^2 +v_P^2 q_0^2) P_{q_0,\Omega/2} -i (\Omega/2) \beta P_{q_0,\Omega/2} = \\ & \qquad \qquad \qquad 2 \gamma Q_{0,\Omega} P_{q_0,-\Omega/2}\label{eq:coupled_minimal_p}
\end{align} 
\resp{We note that generalizing the equations above to a momentum-dependent $\gamma(q)$ amounts to $\gamma\to\gamma(q_0)$ which does not affect the form of the equation or the results.}
We take the resonant limit ($\Omega \to \omega_Q$) and consider  P modes satisfying the parametric resonance condition, which is set by the momentum $q_0$ for which $\delta \omega_P^2 = (\Omega/2)^2 - (\omega_P^2 +v_P^2 q^2)$ vanishes identically. Equations \eqref{eq:coupled_minimal_q}-\eqref{eq:coupled_minimal_p} now become
\begin{align}
    & - i \Omega \beta Q_{0, \Omega} = -E_0/2 + 2\gamma P_{q_0,\Omega/2} P^*_{q_0,-\Omega/2}, \label{eq:res_Q} \\
    & i\beta \Omega/2 P_{q_0, \Omega/2} = -2 \gamma Q_{0,\Omega} P_{q_0,-\Omega/2}.\label{eq:res_pq}
\end{align}
The equations for $P_{\pm q_0,-\Omega/2}$ differ by $\Omega\to-\Omega, Q_{0,\Omega}\to Q^*_{0,\Omega}$ in \eqref{eq:res_pq}, implying that $|P_{q_0,\Omega/2}|=|P_{q_0,-\Omega/2}|$. This allows to introduce $P_{q_0,\pm \Omega/2} = |P_{q_0,\Omega/2}|e^{i\varphi_{\pm\Omega}}$. Solving Eq. \eqref{eq:res_pq} for $Q_{0,\Omega}$ and substituting the result in \eqref{eq:res_Q}, we get an equation on $P_{q_0,\pm \Omega/2}$:  $e^{i(\varphi_{\Omega} - \varphi_{-\Omega})} [\beta^2\Omega^2/4\gamma +2 \gamma |P_{q_0,\Omega/2}|^2] = E_0/2$. A solution exists for $E_0>E_c$, where $\varphi_{\Omega} = \varphi_{-\Omega}$ and
\begin{align}
E_0 &\geq \frac{\beta^2 \Omega^2}{2\gamma}, ~~ |P_{q_0,\Omega/2}| = \frac{1}{2\sqrt{\gamma}}\sqrt{E_0 - \beta^2 \Omega^2/(2\gamma)}.
\label{eq:critical_value}
\end{align}
indicating that this spatiotemporal ordering onsets in a continuous transition in excellent agreement with our numerics
(Fig.~\ref{fig:fig1}(c)); the order parameter $|P_{q_0, \Omega/2}|$ vanishes below a critical value of $E_0$ that depends on the dissipation $\beta$ and the nonlinear coupling $\gamma$. 
Note that the individual phases $\varphi_{\pm \Omega}$ of $P_{q_0,\pm\Omega/2}$ are undetermined by the above equations, indicating that any value can be realized depending on initial conditions, in agreement with our numerical results. This suggests a spontaneously broken spatial translation symmetry of the steady state.
We include a separate discussion on detuned driving in the SM.
\paragraph{Static Spatial Order---.} Nonlinear self-interaction terms in $V_{\textrm{nl}}$ produce additional static modulated ordering, and here we
focus on the leading term in $V_{nl}$, $\alpha_{P} P^3$. For systems with broken mirror symmetries along $P$, such a term is allowed, for example in rhombohedral BaTiO\textsubscript{3}, tetragonal PbTiO\textsubscript{3} and low symmetry LaGaO\textsubscript{3} \cite{angel2005general}. 
For a finite value of $|P_{q_0,\frac{\Omega}{2}}|$, this cubic coupling acts as a source in the time-independent version of   Eq.~\ref{eq:pmode} which becomes $-v_P^2 \partial_x^2 P = -2\gamma PQ - 3 \alpha P^2$. A solution is obtained
iteratively: Fourier transforming and assuming $P_{2q_0}$ is small, we have 
$4 v_P^2 q_0^2 P_{2q_0} = - 3\alpha P_{q_0, \Omega/2}P_{q_0,-\Omega/2}$. The $QP$ term is neglected as it does not contain Fourier components at $2q_0$ to leading order. The solution is then $|P_{2q_0}| = \frac{3\alpha |P_{q_0,\Omega/2}|^2}{4 v_P^2 q_0^2}$. This allows for the prediction that whereas $|P_{q,\Omega/2}| \sim \sqrt{E-E_c}$, $|P_{2q_0, \Omega}| \sim |E-E_c| \sim |P_{q,\Omega/2}|^2$, in agreement with our numerics (Fig.~\ref{fig:fig3}(b)).
\begin{figure}
    \centering
    \includegraphics[width=\columnwidth]{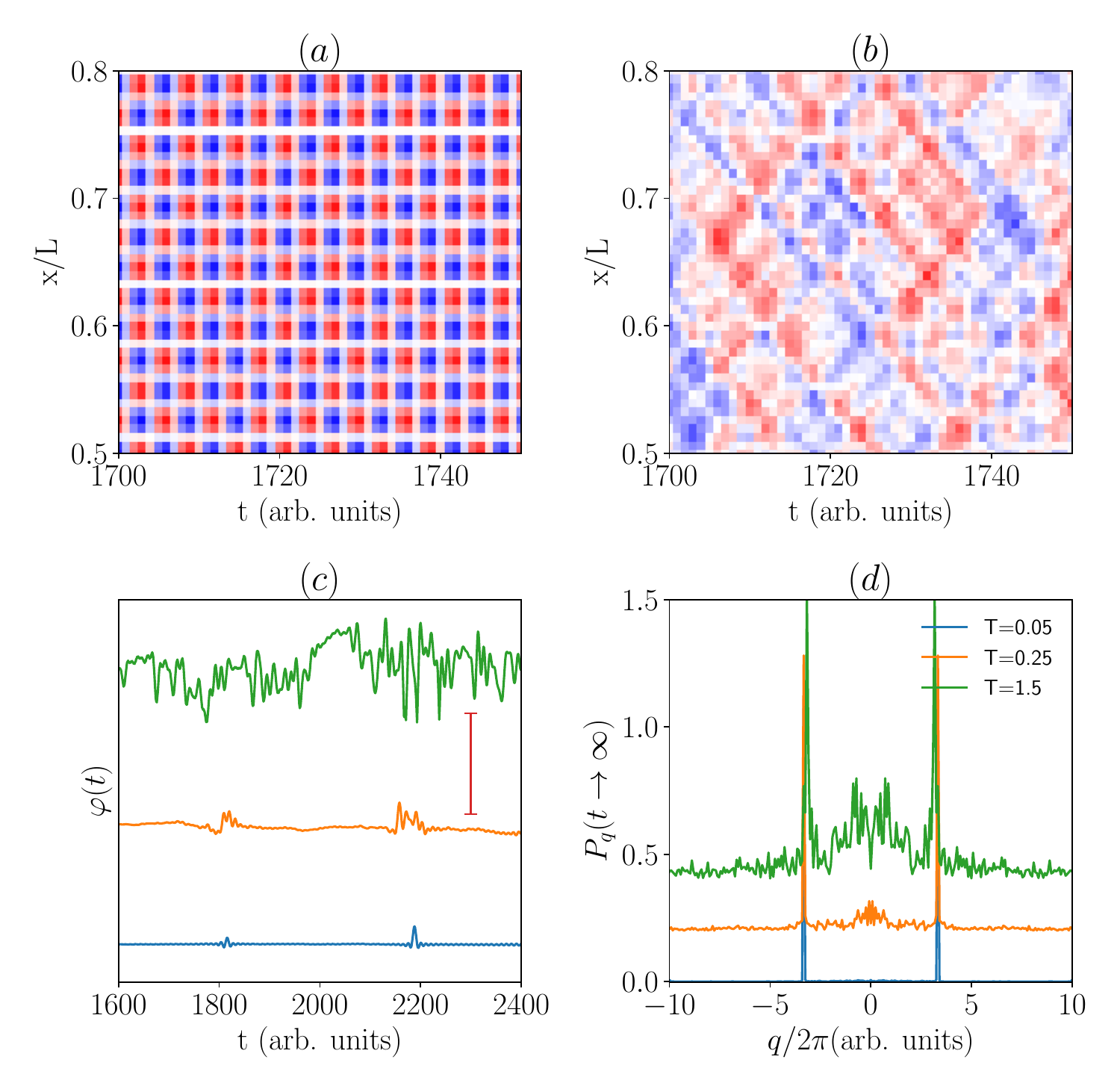}
    \caption{General solution at finite temperature $T \neq 0$. (a) Ordered, fixed phase solution at low temperature, $T = 0.05 \sim \omega_P^2(0) \langle P^2 \rangle$. (b) Substantial phase variation in the solution. Here $T \sim \omega_P^2(q_0)\langle P^2 \rangle$, that is, at the scale of the classical energy of the parametrically excited modes (see main text). (c) Phase of the mode $\varphi$ as a function of $t$ for $\sim 10^{3}$ cycles of the drive. For $T = 0.05$, the phase is nearly perfectly constant in time. For intermediate temperature, e.g. $T = 0.25$, a gradual drift is observed. For temperature on the scale of $\omega_P^2 (q_0) \langle P^2 \rangle$ significant phase fluctuations are observed. Red scale bar measures deviation of $\pi/2$. (d) Momentum decomposition of the condensed mode $P_{q}$ at long times at three temperatures, denoting the low, intermediate and high energy scales (see main text). For all temperatures, the effect allows for thermal occupation of low energy modes, leading to a Gaussian spreading of the weight for all momenta.}
    \label{fig:fig2}
\end{figure}

\paragraph{Finite-Temperature Numerical Results---.}
Above we have shown that the steady state is characterized by a spontaneously broken translation symmetry characterized by a phase $\varphi_{\pm \Omega}$.  In order to assess the stability of this state, we now include temperature modeled as a noise term in our numerical simulations. We find that the phase of $P_{q,
\Omega/2}$
remains unaffected until $T$ is comparable to this mode's classical energy due to driving. In Fig.~\ref{fig:fig2}(a), we show the real-space image of the excited state. For $T$ comparable to the energy at the bottom of $P$ band, $T \sim \omega_p^2 \langle P^2 \rangle$  (where this scale sets the energy of a classical oscillator), the phase remains nearly fixed in time over many cycles of the external drive $\Omega$. When the temperature to meet the scales set by $\omega_P^2 = (\omega_P^2(0) + v_P^2 q_0^2) \langle P^2 \rangle$, significant fluctuations in phase occur and order is ultimately suppressed. At intermediate scales (orange curve in Fig.~\ref{fig:fig2}(c)), the phase exhibits a nearly linear drift, suggesting that it carries out a diffusive random walk as expected for thermal effects. For a generic equilibrium system, such a random walk of the phase is expected to destroy long-range spatial order following the 
Hohenberg-Mermin-Wagner theorem \cite{Mermin66,Halperin19}. Surprisingly, here the phase remains relatively stable for more than $\omega_P t \sim 10^{4}$ cycles for a wide range of $T$ below the classical energy of the mode. The stability of the spatiotemporal order here has been confirmed numerically for a wide range of initial conditions and realizations.
The significant suppression of fluctuations may be attributed to nonlinearities in $P$, as has been observed for models of flocking where nonlinearities stabilize order out of equilibrium 
\cite{Toner95,Toner05,Fruchart21,Diessel22}.
However, our present results can not rule out an alternative scenario, where the system will eventually thermalize after an emergent long prethermalization time \cite{Yao20,Zaletel23}. We leave a detailed investigation of late-time dynamics to future work. As most current experiments on materials are performed with relatively short pump times, rather then continuous pumping, the stability of the order at finite time scales that we demonstrate should be sufficient for the experimental detection of spatiotemporal order.

\paragraph{Generalizations and Material Platforms---.}

Although the results we presented here focused on 1D dispersion of the coupled $P,Q$ modes, they can be easily generalized to 
the 2D and 3D cases. There, a whole manifold of $\vec{q}$ set by 
$|\delta\omega_P(q)| = 0$ would be parametrically excited, though we expect only a discrete number of 
symmetry-related $\vec{q}$ will yield the lowest threshold fluence. Similar behavior has been observed in atomic Bose-Einsten condensates \cite{bukov2015}. 

For polar materials, a key ingredient is a large cubic coupling; such a coupling is allowed for a $P$ mode of any symmetry (including magnetic ones) provided that $Q$ is a polar phonon along the order parameter. 
In the \resp{SM}, we show that the coupling can be generically large if the $P, Q$ originate from the same ion; in this scenario (c.f. Fig.~\ref{fig:schematic}(c)), $\gamma \sim \delta Q/ Q_0$, where $Q_0$ is the ferroelectric order parameter and $\delta Q$ is the drive-induced modulation. 
For a general case, we have estimated $E_c$ for PbTiO\textsubscript{3} taking $E$ phonon at $\approx 275 \textrm{cm}^{-1}$\cite{Choudhury2008} as $Q$, an another E phonon at $\approx 90 \textrm{cm}^{-1}$ as $P$. Using parameters in Ref. \cite{Subedi2015}; we estimate $E_0\approx 3.55 \textrm{MV}/\textrm{cm}$ and $q_0\approx 0.95 \AA^{-1}$, within experimental accessibility. 

We reiterate that our proposal does not require polar order or even inversion symmetry breaking; cubic coupling to the polar mode can be achieved in materials with inversion symmetry via gradient coupling $\delta_{ijkl} Q_i P_j \nabla_k P_l, \delta_{ijkl} \neq 0$ (Eq.~\eqref{eq:cubic_pot}). For a ``frozen" $Q$, such couplings are known as Lifshitz invariants \cite{landau2013statistical} and can be constructed for structural \cite{Stokes1993}, incommensurate \cite{Kopsky_1977}, magnetic \cite{moriya1960anisotropic,dzyaloshinsky1958thermodynamic} and even superconducting \cite{mineev2008,manske2024} orders. 

The magnetic case corresponds to an induced  Dzyaloshinskii-Moriya (DM) interaction \cite{moriya1960anisotropic,dzyaloshinsky1958thermodynamic,son2019unconventional} in a system with appreciable SOC with a dynamical DM vector set by the polar phonon displacement, i.e. $D_{ij}(Q) \mathbf{S}_{i}\times \mathbf{S}_j$, where $D_{ij} \sim Q$. 
In spin wave theory, this term becomes $\propto Q \mathbf{S}(x) \times \frac{\partial \mathbf{S}}{\partial x}$, which is allowed for any symmetry. We leave details of this interaction to the \resp{SM}. 
This idea may open an alternative path for localized magnon generation in the THz regime, allowing for faster nanoscale magnonic computation \cite{Barman_2021}. Intriguingly, the intensely studied NiO, has TO phonons \cite{niophonons} at about $30-40$ meV, while the magnon band extends from few to $100$ meV \cite{niomagnons}, opening the path to the creation of spatiotemporal order with THz magnons. Another potential application of our proposal is in accessing magnon dispersion in 2D materials, such as CrI\textsubscript{3}, where magnons close to the $K$ point are predicted to posses a topological gap \cite{costa2020topological}.

Finally, parametric phenomena have a long history in physics \cite{Landau76} and continue to be explored in a variety of systems \cite{Yao20,Zaletel23,Kleiner21}. The key novelty of our proposal is to use nonlinear phononics to induce strong parametric phenomena in solids. For magnetic materials, a qualitative advantage of the mechanism proposed in this work is that the excitation occurs by electric field, allowing for a much stronger impact on the material than magnetic field, typically used.

\paragraph{Summary and outlook---.}
Here we have presented a mechanism for spatiotemporal ordering in solids using a nonlinear parametric resonance created by a driven optical phonon. The presented mechanism  is a direct way of generating tunable finite momentum order with light. With additional nonlinearities in the excited field $P$, it is possible to produce static order that spatially oscillates at twice the momentum $2q_0$ of the parametrically excited mode. We have also shown remarkable robustness of the parametric excitation to temperature.
We make specific predictions for Bragg diffraction measurements in candidate materials involving both polarization and magnetization spatiotemporal orderings.
\begin{acknowledgements}
\noindent{\it Acknowledgements.}
We thank B. L. Altshuler, K. Behnia, A.V. Balatsky, A. Cavalleri, P. E. Dolgirev, B. Fauqu\'e, 
R. Moessner, A. Sengupta
and S.L. Sondhi for inspiring discussions. D. K. and A.C. are supported by Abrahams Postdoctoral Fellowships of the Center for Materials Theory, Rutgers University and D.K. is also supported by the Zuckerman STEM Fellowship. This research was supported in part by grant NSF PHY-2309135 to the Kavli Institute for Theoretical Physics (KITP) [PAV]. PAV acknowledges support by the University of Connecticut OVPR Quantum CT seed grant.
\end{acknowledgements}
\bibliography{main.bbl}

\begin{thebibliography}{66}%
\makeatletter
\providecommand \@ifxundefined [1]{%
 \@ifx{#1\undefined}
}%
\providecommand \@ifnum [1]{%
 \ifnum #1\expandafter \@firstoftwo
 \else \expandafter \@secondoftwo
 \fi
}%
\providecommand \@ifx [1]{%
 \ifx #1\expandafter \@firstoftwo
 \else \expandafter \@secondoftwo
 \fi
}%
\providecommand \natexlab [1]{#1}%
\providecommand \enquote  [1]{``#1''}%
\providecommand \bibnamefont  [1]{#1}%
\providecommand \bibfnamefont [1]{#1}%
\providecommand \citenamefont [1]{#1}%
\providecommand \href@noop [0]{\@secondoftwo}%
\providecommand \href [0]{\begingroup \@sanitize@url \@href}%
\providecommand \@href[1]{\@@startlink{#1}\@@href}%
\providecommand \@@href[1]{\endgroup#1\@@endlink}%
\providecommand \@sanitize@url [0]{\catcode `\\12\catcode `\$12\catcode
  `\&12\catcode `\#12\catcode `\^12\catcode `\_12\catcode `\%12\relax}%
\providecommand \@@startlink[1]{}%
\providecommand \@@endlink[0]{}%
\providecommand \url  [0]{\begingroup\@sanitize@url \@url }%
\providecommand \@url [1]{\endgroup\@href {#1}{\urlprefix }}%
\providecommand \urlprefix  [0]{URL }%
\providecommand \Eprint [0]{\href }%
\providecommand \doibase [0]{http://dx.doi.org/}%
\providecommand \selectlanguage [0]{\@gobble}%
\providecommand \bibinfo  [0]{\@secondoftwo}%
\providecommand \bibfield  [0]{\@secondoftwo}%
\providecommand \translation [1]{[#1]}%
\providecommand \BibitemOpen [0]{}%
\providecommand \bibitemStop [0]{}%
\providecommand \bibitemNoStop [0]{.\EOS\space}%
\providecommand \EOS [0]{\spacefactor3000\relax}%
\providecommand \BibitemShut  [1]{\csname bibitem#1\endcsname}%
\let\auto@bib@innerbib\@empty
\bibitem [{\citenamefont {Basov}\ \emph {et~al.}(2017)\citenamefont {Basov},
  \citenamefont {Averitt},\ and\ \citenamefont {Hsieh}}]{basov2017}%
  \BibitemOpen
  \bibfield  {author} {\bibinfo {author} {\bibfnamefont {D.}~\bibnamefont
  {Basov}}, \bibinfo {author} {\bibfnamefont {R.}~\bibnamefont {Averitt}}, \
  and\ \bibinfo {author} {\bibfnamefont {D.}~\bibnamefont {Hsieh}},\
  }\href@noop {} {\bibfield  {journal} {\bibinfo  {journal} {Nature materials}\
  }\textbf {\bibinfo {volume} {16}},\ \bibinfo {pages} {1077} (\bibinfo {year}
  {2017})}\BibitemShut {NoStop}%
\bibitem [{\citenamefont {de~la Torre}\ \emph {et~al.}(2021)\citenamefont
  {de~la Torre}, \citenamefont {Kennes}, \citenamefont {Claassen},
  \citenamefont {Gerber}, \citenamefont {McIver},\ and\ \citenamefont
  {Sentef}}]{de2021colloquium}%
  \BibitemOpen
  \bibfield  {author} {\bibinfo {author} {\bibfnamefont {A.}~\bibnamefont
  {de~la Torre}}, \bibinfo {author} {\bibfnamefont {D.~M.}\ \bibnamefont
  {Kennes}}, \bibinfo {author} {\bibfnamefont {M.}~\bibnamefont {Claassen}},
  \bibinfo {author} {\bibfnamefont {S.}~\bibnamefont {Gerber}}, \bibinfo
  {author} {\bibfnamefont {J.~W.}\ \bibnamefont {McIver}}, \ and\ \bibinfo
  {author} {\bibfnamefont {M.~A.}\ \bibnamefont {Sentef}},\ }\href@noop {}
  {\bibfield  {journal} {\bibinfo  {journal} {Reviews of Modern Physics}\
  }\textbf {\bibinfo {volume} {93}},\ \bibinfo {pages} {041002} (\bibinfo
  {year} {2021})}\BibitemShut {NoStop}%
\bibitem [{\citenamefont {Bloch}\ \emph {et~al.}(2022)\citenamefont {Bloch},
  \citenamefont {Cavalleri}, \citenamefont {Galitskii}, \citenamefont
  {Harezi},\ and\ \citenamefont {Rubio}}]{Bloch22}%
  \BibitemOpen
  \bibfield  {author} {\bibinfo {author} {\bibfnamefont {J.}~\bibnamefont
  {Bloch}}, \bibinfo {author} {\bibfnamefont {A.}~\bibnamefont {Cavalleri}},
  \bibinfo {author} {\bibfnamefont {V.}~\bibnamefont {Galitskii}}, \bibinfo
  {author} {\bibfnamefont {M.}~\bibnamefont {Harezi}}, \ and\ \bibinfo {author}
  {\bibfnamefont {A.}~\bibnamefont {Rubio}},\ }\href@noop {} {\bibfield
  {journal} {\bibinfo  {journal} {Nature}\ }\textbf {\bibinfo {volume} {606}},\
  \bibinfo {pages} {41} (\bibinfo {year} {2022})}\BibitemShut {NoStop}%
\bibitem [{\citenamefont {Hebling}\ \emph {et~al.}(2008)\citenamefont
  {Hebling}, \citenamefont {Yeh}, \citenamefont {Hoffmann}, \citenamefont
  {Bartal},\ and\ \citenamefont {Nelson}}]{hebling2008}%
  \BibitemOpen
  \bibfield  {author} {\bibinfo {author} {\bibfnamefont {J.}~\bibnamefont
  {Hebling}}, \bibinfo {author} {\bibfnamefont {K.-L.}\ \bibnamefont {Yeh}},
  \bibinfo {author} {\bibfnamefont {M.~C.}\ \bibnamefont {Hoffmann}}, \bibinfo
  {author} {\bibfnamefont {B.}~\bibnamefont {Bartal}}, \ and\ \bibinfo {author}
  {\bibfnamefont {K.~A.}\ \bibnamefont {Nelson}},\ }\href@noop {} {\bibfield
  {journal} {\bibinfo  {journal} {JOSA B}\ }\textbf {\bibinfo {volume} {25}},\
  \bibinfo {pages} {B6} (\bibinfo {year} {2008})}\BibitemShut {NoStop}%
\bibitem [{\citenamefont {Emma}\ \emph {et~al.}(2010)\citenamefont {Emma},
  \citenamefont {Akre}, \citenamefont {Arthur}, \citenamefont {Bionta},
  \citenamefont {Bostedt}, \citenamefont {Bozek}, \citenamefont {Brachmann},
  \citenamefont {Bucksbaum}, \citenamefont {Coffee}, \citenamefont {Decker}
  \emph {et~al.}}]{emma2010}%
  \BibitemOpen
  \bibfield  {author} {\bibinfo {author} {\bibfnamefont {P.}~\bibnamefont
  {Emma}}, \bibinfo {author} {\bibfnamefont {R.}~\bibnamefont {Akre}}, \bibinfo
  {author} {\bibfnamefont {J.}~\bibnamefont {Arthur}}, \bibinfo {author}
  {\bibfnamefont {R.}~\bibnamefont {Bionta}}, \bibinfo {author} {\bibfnamefont
  {C.}~\bibnamefont {Bostedt}}, \bibinfo {author} {\bibfnamefont
  {J.}~\bibnamefont {Bozek}}, \bibinfo {author} {\bibfnamefont
  {A.}~\bibnamefont {Brachmann}}, \bibinfo {author} {\bibfnamefont
  {P.}~\bibnamefont {Bucksbaum}}, \bibinfo {author} {\bibfnamefont
  {R.}~\bibnamefont {Coffee}}, \bibinfo {author} {\bibfnamefont {F.-J.}\
  \bibnamefont {Decker}},  \emph {et~al.},\ }\href@noop {} {\bibfield
  {journal} {\bibinfo  {journal} {nature photonics}\ }\textbf {\bibinfo
  {volume} {4}},\ \bibinfo {pages} {641} (\bibinfo {year} {2010})}\BibitemShut
  {NoStop}%
\bibitem [{\citenamefont {Buzzi}\ \emph {et~al.}(2018)\citenamefont {Buzzi},
  \citenamefont {F\"orst}, \citenamefont {Mankowsky},\ and\ \citenamefont
  {Cavalleri}}]{Buzzi18}%
  \BibitemOpen
  \bibfield  {author} {\bibinfo {author} {\bibfnamefont {M.}~\bibnamefont
  {Buzzi}}, \bibinfo {author} {\bibfnamefont {M.}~\bibnamefont {F\"orst}},
  \bibinfo {author} {\bibfnamefont {R.}~\bibnamefont {Mankowsky}}, \ and\
  \bibinfo {author} {\bibfnamefont {A.}~\bibnamefont {Cavalleri}},\ }\href@noop
  {} {\bibfield  {journal} {\bibinfo  {journal} {Nature Reviews Materials}\
  }\textbf {\bibinfo {volume} {3}},\ \bibinfo {pages} {299} (\bibinfo {year}
  {2018})}\BibitemShut {NoStop}%
\bibitem [{\citenamefont {Disa}\ \emph {et~al.}(2021)\citenamefont {Disa},
  \citenamefont {Nova},\ and\ \citenamefont {Cavalleri}}]{Disa21}%
  \BibitemOpen
  \bibfield  {author} {\bibinfo {author} {\bibfnamefont {A.}~\bibnamefont
  {Disa}}, \bibinfo {author} {\bibfnamefont {T.}~\bibnamefont {Nova}}, \ and\
  \bibinfo {author} {\bibfnamefont {A.}~\bibnamefont {Cavalleri}},\ }\href@noop
  {} {\bibfield  {journal} {\bibinfo  {journal} {Nature Physics}\ }\textbf
  {\bibinfo {volume} {17}},\ \bibinfo {pages} {1087} (\bibinfo {year}
  {2021})}\BibitemShut {NoStop}%
\bibitem [{\citenamefont {Henstridge}\ \emph {et~al.}(2022)\citenamefont
  {Henstridge}, \citenamefont {F\"orst}, \citenamefont {Rowe}, \citenamefont
  {Fechner},\ and\ \citenamefont {Cavalleri}}]{Henstridge22}%
  \BibitemOpen
  \bibfield  {author} {\bibinfo {author} {\bibfnamefont {M.}~\bibnamefont
  {Henstridge}}, \bibinfo {author} {\bibfnamefont {M.}~\bibnamefont {F\"orst}},
  \bibinfo {author} {\bibfnamefont {E.}~\bibnamefont {Rowe}}, \bibinfo {author}
  {\bibfnamefont {M.}~\bibnamefont {Fechner}}, \ and\ \bibinfo {author}
  {\bibfnamefont {A.}~\bibnamefont {Cavalleri}},\ }\href@noop {} {\bibfield
  {journal} {\bibinfo  {journal} {Nature Physics}\ }\textbf {\bibinfo {volume}
  {18}},\ \bibinfo {pages} {457} (\bibinfo {year} {2022})}\BibitemShut
  {NoStop}%
\bibitem [{\citenamefont {Khemani}\ \emph {et~al.}(2019)\citenamefont
  {Khemani}, \citenamefont {Moessner},\ and\ \citenamefont
  {Sondhi}}]{Khemani19}%
  \BibitemOpen
  \bibfield  {author} {\bibinfo {author} {\bibfnamefont {V.}~\bibnamefont
  {Khemani}}, \bibinfo {author} {\bibfnamefont {R.}~\bibnamefont {Moessner}}, \
  and\ \bibinfo {author} {\bibfnamefont {S.~L.}\ \bibnamefont {Sondhi}},\
  }\href@noop {} {\bibfield  {journal} {\bibinfo  {journal} {arXiv:1910.10745}\
  } (\bibinfo {year} {2019})}\BibitemShut {NoStop}%
\bibitem [{\citenamefont {Zaletel}\ \emph {et~al.}(2023)\citenamefont
  {Zaletel}, \citenamefont {Lukin}, \citenamefont {Monroe}, \citenamefont
  {Nayak}, \citenamefont {Wilczek},\ and\ \citenamefont {Yao}}]{Zaletel23}%
  \BibitemOpen
  \bibfield  {author} {\bibinfo {author} {\bibfnamefont {M.~P.}\ \bibnamefont
  {Zaletel}}, \bibinfo {author} {\bibfnamefont {M.}~\bibnamefont {Lukin}},
  \bibinfo {author} {\bibfnamefont {C.}~\bibnamefont {Monroe}}, \bibinfo
  {author} {\bibfnamefont {C.}~\bibnamefont {Nayak}}, \bibinfo {author}
  {\bibfnamefont {F.}~\bibnamefont {Wilczek}}, \ and\ \bibinfo {author}
  {\bibfnamefont {N.~Y.}\ \bibnamefont {Yao}},\ }\href {\doibase
  10.1103/RevModPhys.95.031001} {\bibfield  {journal} {\bibinfo  {journal}
  {Rev. Mod. Phys.}\ }\textbf {\bibinfo {volume} {95}},\ \bibinfo {pages}
  {031001} (\bibinfo {year} {2023})}\BibitemShut {NoStop}%
\bibitem [{\citenamefont {Shapere}\ and\ \citenamefont
  {Wilczek}(2012)}]{Shapere12}%
  \BibitemOpen
  \bibfield  {author} {\bibinfo {author} {\bibfnamefont {A.}~\bibnamefont
  {Shapere}}\ and\ \bibinfo {author} {\bibfnamefont {F.}~\bibnamefont
  {Wilczek}},\ }\href {\doibase 10.1103/PhysRevLett.109.160402} {\bibfield
  {journal} {\bibinfo  {journal} {Phys. Rev. Lett.}\ }\textbf {\bibinfo
  {volume} {109}},\ \bibinfo {pages} {160402} (\bibinfo {year}
  {2012})}\BibitemShut {NoStop}%
\bibitem [{\citenamefont {Yao}\ \emph {et~al.}(2020)\citenamefont {Yao},
  \citenamefont {Nayak}, \citenamefont {Balents},\ and\ \citenamefont
  {Zalatel}}]{Yao20}%
  \BibitemOpen
  \bibfield  {author} {\bibinfo {author} {\bibfnamefont {N.~Y.}\ \bibnamefont
  {Yao}}, \bibinfo {author} {\bibfnamefont {C.}~\bibnamefont {Nayak}}, \bibinfo
  {author} {\bibfnamefont {L.}~\bibnamefont {Balents}}, \ and\ \bibinfo
  {author} {\bibfnamefont {M.~P.}\ \bibnamefont {Zalatel}},\ }\href@noop {}
  {\bibfield  {journal} {\bibinfo  {journal} {Nature Physics}\ }\textbf
  {\bibinfo {volume} {16}},\ \bibinfo {pages} {438} (\bibinfo {year}
  {2020})}\BibitemShut {NoStop}%
\bibitem [{\citenamefont {Wilczek}(2012)}]{Wilczek12}%
  \BibitemOpen
  \bibfield  {author} {\bibinfo {author} {\bibfnamefont {F.}~\bibnamefont
  {Wilczek}},\ }\href {\doibase 10.1103/PhysRevLett.109.160401} {\bibfield
  {journal} {\bibinfo  {journal} {Phys. Rev. Lett.}\ }\textbf {\bibinfo
  {volume} {109}},\ \bibinfo {pages} {160401} (\bibinfo {year}
  {2012})}\BibitemShut {NoStop}%
\bibitem [{\citenamefont {Khemani}\ \emph {et~al.}(2016)\citenamefont
  {Khemani}, \citenamefont {Lazarides}, \citenamefont {Moessner},\ and\
  \citenamefont {Sondhi}}]{Khemani16}%
  \BibitemOpen
  \bibfield  {author} {\bibinfo {author} {\bibfnamefont {V.}~\bibnamefont
  {Khemani}}, \bibinfo {author} {\bibfnamefont {A.}~\bibnamefont {Lazarides}},
  \bibinfo {author} {\bibfnamefont {R.}~\bibnamefont {Moessner}}, \ and\
  \bibinfo {author} {\bibfnamefont {S.~L.}\ \bibnamefont {Sondhi}},\ }\href
  {\doibase 10.1103/PhysRevLett.116.250401} {\bibfield  {journal} {\bibinfo
  {journal} {Phys. Rev. Lett.}\ }\textbf {\bibinfo {volume} {116}},\ \bibinfo
  {pages} {250401} (\bibinfo {year} {2016})}\BibitemShut {NoStop}%
\bibitem [{\citenamefont {Nova}\ \emph {et~al.}(2019)\citenamefont {Nova},
  \citenamefont {Disa}, \citenamefont {Fechner},\ and\ \citenamefont
  {Cavalleri}}]{nova2019}%
  \BibitemOpen
  \bibfield  {author} {\bibinfo {author} {\bibfnamefont {T.}~\bibnamefont
  {Nova}}, \bibinfo {author} {\bibfnamefont {A.}~\bibnamefont {Disa}}, \bibinfo
  {author} {\bibfnamefont {M.}~\bibnamefont {Fechner}}, \ and\ \bibinfo
  {author} {\bibfnamefont {A.}~\bibnamefont {Cavalleri}},\ }\href@noop {}
  {\bibfield  {journal} {\bibinfo  {journal} {Science}\ }\textbf {\bibinfo
  {volume} {364}},\ \bibinfo {pages} {1075} (\bibinfo {year}
  {2019})}\BibitemShut {NoStop}%
\bibitem [{\citenamefont {Li}\ \emph {et~al.}(2019)\citenamefont {Li},
  \citenamefont {Qiu}, \citenamefont {Zhang}, \citenamefont {Baldini},
  \citenamefont {Lu}, \citenamefont {Rappe},\ and\ \citenamefont
  {Nelson}}]{li2019terahertz}%
  \BibitemOpen
  \bibfield  {author} {\bibinfo {author} {\bibfnamefont {X.}~\bibnamefont
  {Li}}, \bibinfo {author} {\bibfnamefont {T.}~\bibnamefont {Qiu}}, \bibinfo
  {author} {\bibfnamefont {J.}~\bibnamefont {Zhang}}, \bibinfo {author}
  {\bibfnamefont {E.}~\bibnamefont {Baldini}}, \bibinfo {author} {\bibfnamefont
  {J.}~\bibnamefont {Lu}}, \bibinfo {author} {\bibfnamefont {A.~M.}\
  \bibnamefont {Rappe}}, \ and\ \bibinfo {author} {\bibfnamefont {K.~A.}\
  \bibnamefont {Nelson}},\ }\href@noop {} {\bibfield  {journal} {\bibinfo
  {journal} {Science}\ }\textbf {\bibinfo {volume} {364}},\ \bibinfo {pages}
  {1079} (\bibinfo {year} {2019})}\BibitemShut {NoStop}%
\bibitem [{\citenamefont {Disa}\ \emph {et~al.}(2023)\citenamefont {Disa},
  \citenamefont {Curtis}, \citenamefont {Fechner}, \citenamefont {Liu},
  \citenamefont {Von~Hoegen}, \citenamefont {F{\"o}rst}, \citenamefont {Nova},
  \citenamefont {Narang}, \citenamefont {Maljuk}, \citenamefont {Boris} \emph
  {et~al.}}]{disa2023photo}%
  \BibitemOpen
  \bibfield  {author} {\bibinfo {author} {\bibfnamefont {A.}~\bibnamefont
  {Disa}}, \bibinfo {author} {\bibfnamefont {J.}~\bibnamefont {Curtis}},
  \bibinfo {author} {\bibfnamefont {M.}~\bibnamefont {Fechner}}, \bibinfo
  {author} {\bibfnamefont {A.}~\bibnamefont {Liu}}, \bibinfo {author}
  {\bibfnamefont {A.}~\bibnamefont {Von~Hoegen}}, \bibinfo {author}
  {\bibfnamefont {M.}~\bibnamefont {F{\"o}rst}}, \bibinfo {author}
  {\bibfnamefont {T.}~\bibnamefont {Nova}}, \bibinfo {author} {\bibfnamefont
  {P.}~\bibnamefont {Narang}}, \bibinfo {author} {\bibfnamefont
  {A.}~\bibnamefont {Maljuk}}, \bibinfo {author} {\bibfnamefont
  {A.}~\bibnamefont {Boris}},  \emph {et~al.},\ }\href@noop {} {\bibfield
  {journal} {\bibinfo  {journal} {Nature}\ }\textbf {\bibinfo {volume} {617}},\
  \bibinfo {pages} {73} (\bibinfo {year} {2023})}\BibitemShut {NoStop}%
\bibitem [{\citenamefont {Cheng}\ \emph {et~al.}(2023)\citenamefont {Cheng},
  \citenamefont {Kramer}, \citenamefont {Shen},\ and\ \citenamefont
  {Hoffmann}}]{cheng2023}%
  \BibitemOpen
  \bibfield  {author} {\bibinfo {author} {\bibfnamefont {B.}~\bibnamefont
  {Cheng}}, \bibinfo {author} {\bibfnamefont {P.~L.}\ \bibnamefont {Kramer}},
  \bibinfo {author} {\bibfnamefont {Z.-X.}\ \bibnamefont {Shen}}, \ and\
  \bibinfo {author} {\bibfnamefont {M.~C.}\ \bibnamefont {Hoffmann}},\
  }\href@noop {} {\bibfield  {journal} {\bibinfo  {journal} {Physical Review
  Letters}\ }\textbf {\bibinfo {volume} {130}},\ \bibinfo {pages} {126902}
  (\bibinfo {year} {2023})}\BibitemShut {NoStop}%
\bibitem [{\citenamefont {{Fechner}}\ \emph {et~al.}(2024)\citenamefont
  {{Fechner}}, \citenamefont {{F{\"o}rst}}, \citenamefont {{Orenstein}},
  \citenamefont {{Krapivin}}, \citenamefont {{Disa}}, \citenamefont {{Buzzi}},
  \citenamefont {{von Hoegen}}, \citenamefont {{de la Pena}}, \citenamefont
  {{Nguyen}}, \citenamefont {{Mankowsky}}, \citenamefont {{Sander}},
  \citenamefont {{Lemke}}, \citenamefont {{Deng}}, \citenamefont {{Trigo}},\
  and\ \citenamefont {{Cavalleri}}}]{Fechner2024}%
  \BibitemOpen
  \bibfield  {author} {\bibinfo {author} {\bibfnamefont {M.}~\bibnamefont
  {{Fechner}}}, \bibinfo {author} {\bibfnamefont {M.}~\bibnamefont
  {{F{\"o}rst}}}, \bibinfo {author} {\bibfnamefont {G.}~\bibnamefont
  {{Orenstein}}}, \bibinfo {author} {\bibfnamefont {V.}~\bibnamefont
  {{Krapivin}}}, \bibinfo {author} {\bibfnamefont {A.~S.}\ \bibnamefont
  {{Disa}}}, \bibinfo {author} {\bibfnamefont {M.}~\bibnamefont {{Buzzi}}},
  \bibinfo {author} {\bibfnamefont {A.}~\bibnamefont {{von Hoegen}}}, \bibinfo
  {author} {\bibfnamefont {G.}~\bibnamefont {{de la Pena}}}, \bibinfo {author}
  {\bibfnamefont {Q.~L.}\ \bibnamefont {{Nguyen}}}, \bibinfo {author}
  {\bibfnamefont {R.}~\bibnamefont {{Mankowsky}}}, \bibinfo {author}
  {\bibfnamefont {M.}~\bibnamefont {{Sander}}}, \bibinfo {author}
  {\bibfnamefont {H.}~\bibnamefont {{Lemke}}}, \bibinfo {author} {\bibfnamefont
  {Y.}~\bibnamefont {{Deng}}}, \bibinfo {author} {\bibfnamefont
  {M.}~\bibnamefont {{Trigo}}}, \ and\ \bibinfo {author} {\bibfnamefont
  {A.}~\bibnamefont {{Cavalleri}}},\ }\href {\doibase
  10.1038/s41563-023-01791-y} {\bibfield  {journal} {\bibinfo  {journal}
  {Nature Materials}\ }\textbf {\bibinfo {volume} {23}},\ \bibinfo {pages}
  {363} (\bibinfo {year} {2024})},\ \Eprint {http://arxiv.org/abs/2301.08703}
  {arXiv:2301.08703 [cond-mat.mtrl-sci]} \BibitemShut {NoStop}%
\bibitem [{\citenamefont {Orenstein}\ \emph {et~al.}(2024)\citenamefont
  {Orenstein}, \citenamefont {Krapivin}, \citenamefont {Huang}, \citenamefont
  {Zhang}, \citenamefont {de~la Pena~Munoz}, \citenamefont {Duncan},
  \citenamefont {Nguyen}, \citenamefont {Stanton}, \citenamefont {Teitelbaum},
  \citenamefont {Yavas}, \citenamefont {Sato}, \citenamefont {Hoffman},
  \citenamefont {Kramer}, \citenamefont {Zhang}, \citenamefont {Cavalleri},
  \citenamefont {Comin}, \citenamefont {Dean}, \citenamefont {Disa},
  \citenamefont {F\"orst}, \citenamefont {Johnson}, \citenamefont {Mittrano},
  \citenamefont {Rappe}, \citenamefont {Reis}, \citenamefont {Zhu},
  \citenamefont {Nelson},\ and\ \citenamefont {Trigo}}]{Orenstein24}%
  \BibitemOpen
  \bibfield  {author} {\bibinfo {author} {\bibfnamefont {G.}~\bibnamefont
  {Orenstein}}, \bibinfo {author} {\bibfnamefont {V.}~\bibnamefont {Krapivin}},
  \bibinfo {author} {\bibfnamefont {Y.}~\bibnamefont {Huang}}, \bibinfo
  {author} {\bibfnamefont {Z.}~\bibnamefont {Zhang}}, \bibinfo {author}
  {\bibfnamefont {G.}~\bibnamefont {de~la Pena~Munoz}}, \bibinfo {author}
  {\bibfnamefont {R.}~\bibnamefont {Duncan}}, \bibinfo {author} {\bibfnamefont
  {Q.}~\bibnamefont {Nguyen}}, \bibinfo {author} {\bibfnamefont
  {J.}~\bibnamefont {Stanton}}, \bibinfo {author} {\bibfnamefont
  {S.}~\bibnamefont {Teitelbaum}}, \bibinfo {author} {\bibfnamefont
  {H.}~\bibnamefont {Yavas}}, \bibinfo {author} {\bibfnamefont
  {T.}~\bibnamefont {Sato}}, \bibinfo {author} {\bibfnamefont {M.}~\bibnamefont
  {Hoffman}}, \bibinfo {author} {\bibfnamefont {P.}~\bibnamefont {Kramer}},
  \bibinfo {author} {\bibfnamefont {J.}~\bibnamefont {Zhang}}, \bibinfo
  {author} {\bibfnamefont {A.}~\bibnamefont {Cavalleri}}, \bibinfo {author}
  {\bibfnamefont {R.}~\bibnamefont {Comin}}, \bibinfo {author} {\bibfnamefont
  {M.}~\bibnamefont {Dean}}, \bibinfo {author} {\bibfnamefont {A.}~\bibnamefont
  {Disa}}, \bibinfo {author} {\bibfnamefont {M.}~\bibnamefont {F\"orst}},
  \bibinfo {author} {\bibfnamefont {S.}~\bibnamefont {Johnson}}, \bibinfo
  {author} {\bibfnamefont {M.}~\bibnamefont {Mittrano}}, \bibinfo {author}
  {\bibfnamefont {A.}~\bibnamefont {Rappe}}, \bibinfo {author} {\bibfnamefont
  {D.}~\bibnamefont {Reis}}, \bibinfo {author} {\bibfnamefont {D.}~\bibnamefont
  {Zhu}}, \bibinfo {author} {\bibfnamefont {K.}~\bibnamefont {Nelson}}, \ and\
  \bibinfo {author} {\bibfnamefont {M.}~\bibnamefont {Trigo}},\ }\href@noop {}
  {\bibfield  {journal} {\bibinfo  {journal} {arXiv:2403.17203}\ } (\bibinfo
  {year} {2024})}\BibitemShut {NoStop}%
\bibitem [{\citenamefont {Subedi}\ \emph {et~al.}(2014)\citenamefont {Subedi},
  \citenamefont {Cavalleri},\ and\ \citenamefont {Georges}}]{subedi2014}%
  \BibitemOpen
  \bibfield  {author} {\bibinfo {author} {\bibfnamefont {A.}~\bibnamefont
  {Subedi}}, \bibinfo {author} {\bibfnamefont {A.}~\bibnamefont {Cavalleri}}, \
  and\ \bibinfo {author} {\bibfnamefont {A.}~\bibnamefont {Georges}},\ }\href
  {\doibase 10.1103/PhysRevB.89.220301} {\bibfield  {journal} {\bibinfo
  {journal} {Phys. Rev. B}\ }\textbf {\bibinfo {volume} {89}},\ \bibinfo
  {pages} {220301} (\bibinfo {year} {2014})}\BibitemShut {NoStop}%
\bibitem [{\citenamefont {Subedi}(2015)}]{Subedi2015}%
  \BibitemOpen
  \bibfield  {author} {\bibinfo {author} {\bibfnamefont {A.}~\bibnamefont
  {Subedi}},\ }\href {\doibase 10.1103/PhysRevB.92.214303} {\bibfield
  {journal} {\bibinfo  {journal} {Phys. Rev. B}\ }\textbf {\bibinfo {volume}
  {92}},\ \bibinfo {pages} {214303} (\bibinfo {year} {2015})}\BibitemShut
  {NoStop}%
\bibitem [{\citenamefont {Subedi}(2017)}]{Subedi2017}%
  \BibitemOpen
  \bibfield  {author} {\bibinfo {author} {\bibfnamefont {A.}~\bibnamefont
  {Subedi}},\ }\href {\doibase 10.1103/PhysRevB.95.134113} {\bibfield
  {journal} {\bibinfo  {journal} {Phys. Rev. B}\ }\textbf {\bibinfo {volume}
  {95}},\ \bibinfo {pages} {134113} (\bibinfo {year} {2017})}\BibitemShut
  {NoStop}%
\bibitem [{\citenamefont {Zhuang}\ \emph {et~al.}(2023)\citenamefont {Zhuang},
  \citenamefont {Chakraborty}, \citenamefont {Chandra}, \citenamefont
  {Coleman},\ and\ \citenamefont {Volkov}}]{Zhuang23}%
  \BibitemOpen
  \bibfield  {author} {\bibinfo {author} {\bibfnamefont {Z.}~\bibnamefont
  {Zhuang}}, \bibinfo {author} {\bibfnamefont {A.}~\bibnamefont {Chakraborty}},
  \bibinfo {author} {\bibfnamefont {P.}~\bibnamefont {Chandra}}, \bibinfo
  {author} {\bibfnamefont {P.}~\bibnamefont {Coleman}}, \ and\ \bibinfo
  {author} {\bibfnamefont {P.~A.}\ \bibnamefont {Volkov}},\ }\href {\doibase
  10.1103/PhysRevB.107.224307} {\bibfield  {journal} {\bibinfo  {journal}
  {Phys. Rev. B}\ }\textbf {\bibinfo {volume} {107}},\ \bibinfo {pages}
  {224307} (\bibinfo {year} {2023})}\BibitemShut {NoStop}%
\bibitem [{\citenamefont {Puviani}\ and\ \citenamefont
  {Sentef}(2018)}]{Puviani2018}%
  \BibitemOpen
  \bibfield  {author} {\bibinfo {author} {\bibfnamefont {M.}~\bibnamefont
  {Puviani}}\ and\ \bibinfo {author} {\bibfnamefont {M.~A.}\ \bibnamefont
  {Sentef}},\ }\href {\doibase 10.1103/PhysRevB.98.165138} {\bibfield
  {journal} {\bibinfo  {journal} {Phys. Rev. B}\ }\textbf {\bibinfo {volume}
  {98}},\ \bibinfo {pages} {165138} (\bibinfo {year} {2018})}\BibitemShut
  {NoStop}%
\bibitem [{\citenamefont {Radaelli}(2018)}]{Radaelli18}%
  \BibitemOpen
  \bibfield  {author} {\bibinfo {author} {\bibfnamefont {P.~G.}\ \bibnamefont
  {Radaelli}},\ }\href {\doibase 10.1103/PhysRevB.97.085145} {\bibfield
  {journal} {\bibinfo  {journal} {Phys. Rev. B}\ }\textbf {\bibinfo {volume}
  {97}},\ \bibinfo {pages} {085145} (\bibinfo {year} {2018})}\BibitemShut
  {NoStop}%
\bibitem [{\citenamefont {Shin}\ \emph {et~al.}(2022)\citenamefont {Shin},
  \citenamefont {Latini}, \citenamefont {Sch{\"a}fer}, \citenamefont {Sato},
  \citenamefont {Baldini}, \citenamefont {De~Giovannini}, \citenamefont
  {H{\"u}bener},\ and\ \citenamefont {Rubio}}]{rubio2022}%
  \BibitemOpen
  \bibfield  {author} {\bibinfo {author} {\bibfnamefont {D.}~\bibnamefont
  {Shin}}, \bibinfo {author} {\bibfnamefont {S.}~\bibnamefont {Latini}},
  \bibinfo {author} {\bibfnamefont {C.}~\bibnamefont {Sch{\"a}fer}}, \bibinfo
  {author} {\bibfnamefont {S.~A.}\ \bibnamefont {Sato}}, \bibinfo {author}
  {\bibfnamefont {E.}~\bibnamefont {Baldini}}, \bibinfo {author} {\bibfnamefont
  {U.}~\bibnamefont {De~Giovannini}}, \bibinfo {author} {\bibfnamefont
  {H.}~\bibnamefont {H{\"u}bener}}, \ and\ \bibinfo {author} {\bibfnamefont
  {A.}~\bibnamefont {Rubio}},\ }\href@noop {} {\bibfield  {journal} {\bibinfo
  {journal} {Physical Review Letters}\ }\textbf {\bibinfo {volume} {129}},\
  \bibinfo {pages} {167401} (\bibinfo {year} {2022})}\BibitemShut {NoStop}%
\bibitem [{\citenamefont {G\'omez~Pueyo}\ and\ \citenamefont
  {Subedi}(2022)}]{Pueyo22}%
  \BibitemOpen
  \bibfield  {author} {\bibinfo {author} {\bibfnamefont {A.}~\bibnamefont
  {G\'omez~Pueyo}}\ and\ \bibinfo {author} {\bibfnamefont {A.}~\bibnamefont
  {Subedi}},\ }\href {\doibase 10.1103/PhysRevB.106.214305} {\bibfield
  {journal} {\bibinfo  {journal} {Phys. Rev. B}\ }\textbf {\bibinfo {volume}
  {106}},\ \bibinfo {pages} {214305} (\bibinfo {year} {2022})}\BibitemShut
  {NoStop}%
\bibitem [{\citenamefont {Pueyo}\ and\ \citenamefont {Subedi}(2023)}]{Pueyo23}%
  \BibitemOpen
  \bibfield  {author} {\bibinfo {author} {\bibfnamefont {A.~G.}\ \bibnamefont
  {Pueyo}}\ and\ \bibinfo {author} {\bibfnamefont {A.}~\bibnamefont {Subedi}},\
  }\href {\doibase 10.1103/PhysRevB.108.064302} {\bibfield  {journal} {\bibinfo
   {journal} {Phys. Rev. B}\ }\textbf {\bibinfo {volume} {108}},\ \bibinfo
  {pages} {064302} (\bibinfo {year} {2023})}\BibitemShut {NoStop}%
\bibitem [{\citenamefont {Kuzmanovski}\ \emph {et~al.}(2024)\citenamefont
  {Kuzmanovski}, \citenamefont {Schmidt}, \citenamefont {Spaldin},
  \citenamefont {R\o{}nnow}, \citenamefont {Aeppli},\ and\ \citenamefont
  {Balatsky}}]{Kuzmanovskii24}%
  \BibitemOpen
  \bibfield  {author} {\bibinfo {author} {\bibfnamefont {D.}~\bibnamefont
  {Kuzmanovski}}, \bibinfo {author} {\bibfnamefont {J.}~\bibnamefont
  {Schmidt}}, \bibinfo {author} {\bibfnamefont {N.~A.}\ \bibnamefont
  {Spaldin}}, \bibinfo {author} {\bibfnamefont {H.~M.}\ \bibnamefont
  {R\o{}nnow}}, \bibinfo {author} {\bibfnamefont {G.}~\bibnamefont {Aeppli}}, \
  and\ \bibinfo {author} {\bibfnamefont {A.~V.}\ \bibnamefont {Balatsky}},\
  }\href {\doibase 10.1103/PhysRevX.14.021016} {\bibfield  {journal} {\bibinfo
  {journal} {Phys. Rev. X}\ }\textbf {\bibinfo {volume} {14}},\ \bibinfo
  {pages} {021016} (\bibinfo {year} {2024})}\BibitemShut {NoStop}%
\bibitem [{\citenamefont {Khalsa}\ \emph {et~al.}(2024)\citenamefont {Khalsa},
  \citenamefont {Kaaret},\ and\ \citenamefont {Benedek}}]{Khalsa24}%
  \BibitemOpen
  \bibfield  {author} {\bibinfo {author} {\bibfnamefont {G.}~\bibnamefont
  {Khalsa}}, \bibinfo {author} {\bibfnamefont {J.~Z.}\ \bibnamefont {Kaaret}},
  \ and\ \bibinfo {author} {\bibfnamefont {N.~A.}\ \bibnamefont {Benedek}},\
  }\href {\doibase 10.1103/PhysRevB.109.024110} {\bibfield  {journal} {\bibinfo
   {journal} {Phys. Rev. B}\ }\textbf {\bibinfo {volume} {109}},\ \bibinfo
  {pages} {024110} (\bibinfo {year} {2024})}\BibitemShut {NoStop}%
\bibitem [{\citenamefont {de~la Pe{\~n}a~Mu{\~n}oz}\ \emph
  {et~al.}(2023)\citenamefont {de~la Pe{\~n}a~Mu{\~n}oz}, \citenamefont
  {Correa}, \citenamefont {Yang}, \citenamefont {Delaire}, \citenamefont
  {Huang}, \citenamefont {Johnson}, \citenamefont {Katayama}, \citenamefont
  {Krapivin}, \citenamefont {Pastor}, \citenamefont {Reis} \emph
  {et~al.}}]{de2023ultrafast}%
  \BibitemOpen
  \bibfield  {author} {\bibinfo {author} {\bibfnamefont {G.~A.}\ \bibnamefont
  {de~la Pe{\~n}a~Mu{\~n}oz}}, \bibinfo {author} {\bibfnamefont {A.~A.}\
  \bibnamefont {Correa}}, \bibinfo {author} {\bibfnamefont {S.}~\bibnamefont
  {Yang}}, \bibinfo {author} {\bibfnamefont {O.}~\bibnamefont {Delaire}},
  \bibinfo {author} {\bibfnamefont {Y.}~\bibnamefont {Huang}}, \bibinfo
  {author} {\bibfnamefont {A.~S.}\ \bibnamefont {Johnson}}, \bibinfo {author}
  {\bibfnamefont {T.}~\bibnamefont {Katayama}}, \bibinfo {author}
  {\bibfnamefont {V.}~\bibnamefont {Krapivin}}, \bibinfo {author}
  {\bibfnamefont {E.}~\bibnamefont {Pastor}}, \bibinfo {author} {\bibfnamefont
  {D.~A.}\ \bibnamefont {Reis}},  \emph {et~al.},\ }\href@noop {} {\bibfield
  {journal} {\bibinfo  {journal} {Nature Physics}\ }\textbf {\bibinfo {volume}
  {19}},\ \bibinfo {pages} {1489} (\bibinfo {year} {2023})}\BibitemShut
  {NoStop}%
\bibitem [{\citenamefont {Huang}\ \emph {et~al.}(2024)\citenamefont {Huang},
  \citenamefont {Sun}, \citenamefont {Teitelbaum}, \citenamefont {Li},
  \citenamefont {Sun}, \citenamefont {Wang}, \citenamefont {Song},
  \citenamefont {Sato}, \citenamefont {Chollet}, \citenamefont {Osaka},
  \citenamefont {Inoue}, \citenamefont {Duncan}, \citenamefont {Shin},
  \citenamefont {Haber}, \citenamefont {Zhou}, \citenamefont {Bernardi},
  \citenamefont {Gu}, \citenamefont {Rondinelli}, \citenamefont {Trigo},
  \citenamefont {Yabashi}, \citenamefont {Maznev}, \citenamefont {Nelson},
  \citenamefont {Zhu},\ and\ \citenamefont {Reis}}]{huang2024hard}%
  \BibitemOpen
  \bibfield  {author} {\bibinfo {author} {\bibfnamefont {Y.}~\bibnamefont
  {Huang}}, \bibinfo {author} {\bibfnamefont {P.}~\bibnamefont {Sun}}, \bibinfo
  {author} {\bibfnamefont {S.~W.}\ \bibnamefont {Teitelbaum}}, \bibinfo
  {author} {\bibfnamefont {H.}~\bibnamefont {Li}}, \bibinfo {author}
  {\bibfnamefont {Y.}~\bibnamefont {Sun}}, \bibinfo {author} {\bibfnamefont
  {N.}~\bibnamefont {Wang}}, \bibinfo {author} {\bibfnamefont {S.}~\bibnamefont
  {Song}}, \bibinfo {author} {\bibfnamefont {T.}~\bibnamefont {Sato}}, \bibinfo
  {author} {\bibfnamefont {M.}~\bibnamefont {Chollet}}, \bibinfo {author}
  {\bibfnamefont {T.}~\bibnamefont {Osaka}}, \bibinfo {author} {\bibfnamefont
  {I.}~\bibnamefont {Inoue}}, \bibinfo {author} {\bibfnamefont {R.~A.}\
  \bibnamefont {Duncan}}, \bibinfo {author} {\bibfnamefont {H.~D.}\
  \bibnamefont {Shin}}, \bibinfo {author} {\bibfnamefont {J.}~\bibnamefont
  {Haber}}, \bibinfo {author} {\bibfnamefont {J.}~\bibnamefont {Zhou}},
  \bibinfo {author} {\bibfnamefont {M.}~\bibnamefont {Bernardi}}, \bibinfo
  {author} {\bibfnamefont {M.}~\bibnamefont {Gu}}, \bibinfo {author}
  {\bibfnamefont {J.~M.}\ \bibnamefont {Rondinelli}}, \bibinfo {author}
  {\bibfnamefont {M.}~\bibnamefont {Trigo}}, \bibinfo {author} {\bibfnamefont
  {M.}~\bibnamefont {Yabashi}}, \bibinfo {author} {\bibfnamefont {A.~A.}\
  \bibnamefont {Maznev}}, \bibinfo {author} {\bibfnamefont {K.~A.}\
  \bibnamefont {Nelson}}, \bibinfo {author} {\bibfnamefont {D.}~\bibnamefont
  {Zhu}}, \ and\ \bibinfo {author} {\bibfnamefont {D.~A.}\ \bibnamefont
  {Reis}},\ }\href@noop {} {\enquote {\bibinfo {title} {Hard x-ray generation
  and detection of nanometer-scale localized coherent acoustic wave packets in
  srtio$_3$ and ktao$_3$},}\ } (\bibinfo {year} {2024}),\ \Eprint
  {http://arxiv.org/abs/2312.16453} {arXiv:2312.16453 [cond-mat.mtrl-sci]}
  \BibitemShut {NoStop}%
\bibitem [{\citenamefont {Fauqu\'e}\ \emph {et~al.}(2022)\citenamefont
  {Fauqu\'e}, \citenamefont {Bourges}, \citenamefont {Subedi}, \citenamefont
  {Behnia}, \citenamefont {Baptiste}, \citenamefont {Roessli}, \citenamefont
  {Fennell}, \citenamefont {Raymond},\ and\ \citenamefont
  {Steffens}}]{Fauque22}%
  \BibitemOpen
  \bibfield  {author} {\bibinfo {author} {\bibfnamefont {B.}~\bibnamefont
  {Fauqu\'e}}, \bibinfo {author} {\bibfnamefont {P.}~\bibnamefont {Bourges}},
  \bibinfo {author} {\bibfnamefont {A.}~\bibnamefont {Subedi}}, \bibinfo
  {author} {\bibfnamefont {K.}~\bibnamefont {Behnia}}, \bibinfo {author}
  {\bibfnamefont {B.}~\bibnamefont {Baptiste}}, \bibinfo {author}
  {\bibfnamefont {B.}~\bibnamefont {Roessli}}, \bibinfo {author} {\bibfnamefont
  {T.}~\bibnamefont {Fennell}}, \bibinfo {author} {\bibfnamefont
  {S.}~\bibnamefont {Raymond}}, \ and\ \bibinfo {author} {\bibfnamefont
  {P.}~\bibnamefont {Steffens}},\ }\href {\doibase
  10.1103/PhysRevB.106.L140301} {\bibfield  {journal} {\bibinfo  {journal}
  {Phys. Rev. B}\ }\textbf {\bibinfo {volume} {106}},\ \bibinfo {pages}
  {L140301} (\bibinfo {year} {2022})}\BibitemShut {NoStop}%
\bibitem [{\citenamefont {Johnson}\ \emph {et~al.}(2023)\citenamefont
  {Johnson}, \citenamefont {Perez-Salinas}, \citenamefont {Siddiqui},
  \citenamefont {Kim}, \citenamefont {Choi}, \citenamefont {Volckaert},
  \citenamefont {Majchrzak}, \citenamefont {Ulstrup}, \citenamefont {Agarwal},
  \citenamefont {Hallman} \emph {et~al.}}]{johnson2023ultrafast}%
  \BibitemOpen
  \bibfield  {author} {\bibinfo {author} {\bibfnamefont {A.~S.}\ \bibnamefont
  {Johnson}}, \bibinfo {author} {\bibfnamefont {D.}~\bibnamefont
  {Perez-Salinas}}, \bibinfo {author} {\bibfnamefont {K.~M.}\ \bibnamefont
  {Siddiqui}}, \bibinfo {author} {\bibfnamefont {S.}~\bibnamefont {Kim}},
  \bibinfo {author} {\bibfnamefont {S.}~\bibnamefont {Choi}}, \bibinfo {author}
  {\bibfnamefont {K.}~\bibnamefont {Volckaert}}, \bibinfo {author}
  {\bibfnamefont {P.~E.}\ \bibnamefont {Majchrzak}}, \bibinfo {author}
  {\bibfnamefont {S.}~\bibnamefont {Ulstrup}}, \bibinfo {author} {\bibfnamefont
  {N.}~\bibnamefont {Agarwal}}, \bibinfo {author} {\bibfnamefont
  {K.}~\bibnamefont {Hallman}},  \emph {et~al.},\ }\href@noop {} {\bibfield
  {journal} {\bibinfo  {journal} {Nature Physics}\ }\textbf {\bibinfo {volume}
  {19}},\ \bibinfo {pages} {215} (\bibinfo {year} {2023})}\BibitemShut
  {NoStop}%
\bibitem [{\citenamefont {Subedi}(2021)}]{subedi2021light}%
  \BibitemOpen
  \bibfield  {author} {\bibinfo {author} {\bibfnamefont {A.}~\bibnamefont
  {Subedi}},\ }\href@noop {} {\bibfield  {journal} {\bibinfo  {journal}
  {Comptes Rendus. Physique}\ }\textbf {\bibinfo {volume} {22}},\ \bibinfo
  {pages} {161} (\bibinfo {year} {2021})}\BibitemShut {NoStop}%
\bibitem [{Note1()}]{Note1}%
  \BibitemOpen
  \bibinfo {note} {See Supplementary material for symmetry analysis, details of
  analytical and numerical calculations and discussion of physical units, which
  includes Refs.~\cite
  {burns1973lattice,tchernyshyov2024field,dormand1980family,Basini2024}}\BibitemShut
  {NoStop}%
\bibitem [{\citenamefont {Choudhury}\ \emph {et~al.}(2008)\citenamefont
  {Choudhury}, \citenamefont {Walter}, \citenamefont {Kolesnikov},\ and\
  \citenamefont {Loong}}]{Choudhury2008}%
  \BibitemOpen
  \bibfield  {author} {\bibinfo {author} {\bibfnamefont {N.}~\bibnamefont
  {Choudhury}}, \bibinfo {author} {\bibfnamefont {E.~J.}\ \bibnamefont
  {Walter}}, \bibinfo {author} {\bibfnamefont {A.~I.}\ \bibnamefont
  {Kolesnikov}}, \ and\ \bibinfo {author} {\bibfnamefont {C.-K.}\ \bibnamefont
  {Loong}},\ }\href {\doibase 10.1103/PhysRevB.77.134111} {\bibfield  {journal}
  {\bibinfo  {journal} {Phys. Rev. B}\ }\textbf {\bibinfo {volume} {77}},\
  \bibinfo {pages} {134111} (\bibinfo {year} {2008})}\BibitemShut {NoStop}%
\bibitem [{\citenamefont {{Leimkuhler}}\ and\ \citenamefont
  {{Matthews}}(2013)}]{leimkuhler2013robust}%
  \BibitemOpen
  \bibfield  {author} {\bibinfo {author} {\bibfnamefont {B.}~\bibnamefont
  {{Leimkuhler}}}\ and\ \bibinfo {author} {\bibfnamefont {C.}~\bibnamefont
  {{Matthews}}},\ }\href {\doibase 10.1063/1.4802990} {\bibfield  {journal}
  {\bibinfo  {journal} {\jcp}\ }\textbf {\bibinfo {volume} {138}},\ \bibinfo
  {eid} {174102} (\bibinfo {year} {2013})},\ \Eprint
  {http://arxiv.org/abs/1304.3269} {arXiv:1304.3269 [physics.comp-ph]}
  \BibitemShut {NoStop}%
\bibitem [{\citenamefont {Lax}(2013)}]{lax2013stability}%
  \BibitemOpen
  \bibfield  {author} {\bibinfo {author} {\bibfnamefont {P.~D.}\ \bibnamefont
  {Lax}},\ }\href@noop {} {\bibfield  {journal} {\bibinfo  {journal} {The
  Courant--Friedrichs--Lewy (CFL) Condition: 80 Years After Its Discovery}\ ,\
  \bibinfo {pages} {1}} (\bibinfo {year} {2013})}\BibitemShut {NoStop}%
\bibitem [{\citenamefont {Landau}\ and\ \citenamefont
  {Lifshitz}(1976)}]{Landau76}%
  \BibitemOpen
  \bibfield  {author} {\bibinfo {author} {\bibfnamefont {L.}~\bibnamefont
  {Landau}}\ and\ \bibinfo {author} {\bibfnamefont {E.}~\bibnamefont
  {Lifshitz}},\ }\href@noop {} {\emph {\bibinfo {title} {Mechanics}}}\
  (\bibinfo  {publisher} {Pergamon Press},\ \bibinfo {year} {1976})\BibitemShut
  {NoStop}%
\bibitem [{\citenamefont {Angel}\ \emph {et~al.}(2005)\citenamefont {Angel},
  \citenamefont {Zhao},\ and\ \citenamefont {Ross}}]{angel2005general}%
  \BibitemOpen
  \bibfield  {author} {\bibinfo {author} {\bibfnamefont {R.~J.}\ \bibnamefont
  {Angel}}, \bibinfo {author} {\bibfnamefont {J.}~\bibnamefont {Zhao}}, \ and\
  \bibinfo {author} {\bibfnamefont {N.~L.}\ \bibnamefont {Ross}},\ }\href@noop
  {} {\bibfield  {journal} {\bibinfo  {journal} {Physical review letters}\
  }\textbf {\bibinfo {volume} {95}},\ \bibinfo {pages} {025503} (\bibinfo
  {year} {2005})}\BibitemShut {NoStop}%
\bibitem [{\citenamefont {Mermin}\ and\ \citenamefont
  {Wagner}(1966)}]{Mermin66}%
  \BibitemOpen
  \bibfield  {author} {\bibinfo {author} {\bibfnamefont {N.~D.}\ \bibnamefont
  {Mermin}}\ and\ \bibinfo {author} {\bibfnamefont {H.}~\bibnamefont
  {Wagner}},\ }\href {\doibase 10.1103/PhysRevLett.17.1133} {\bibfield
  {journal} {\bibinfo  {journal} {Phys. Rev. Lett.}\ }\textbf {\bibinfo
  {volume} {17}},\ \bibinfo {pages} {1133} (\bibinfo {year}
  {1966})}\BibitemShut {NoStop}%
\bibitem [{\citenamefont {Halperin}(2019)}]{Halperin19}%
  \BibitemOpen
  \bibfield  {author} {\bibinfo {author} {\bibfnamefont {B.}~\bibnamefont
  {Halperin}},\ }\href@noop {} {\bibfield  {journal} {\bibinfo  {journal} {J.
  Stat. Phys.}\ }\textbf {\bibinfo {volume} {175}},\ \bibinfo {pages} {521}
  (\bibinfo {year} {2019})}\BibitemShut {NoStop}%
\bibitem [{\citenamefont {Toner}\ and\ \citenamefont {Tu}(1995)}]{Toner95}%
  \BibitemOpen
  \bibfield  {author} {\bibinfo {author} {\bibfnamefont {J.}~\bibnamefont
  {Toner}}\ and\ \bibinfo {author} {\bibfnamefont {Y.}~\bibnamefont {Tu}},\
  }\href {\doibase 10.1103/PhysRevLett.75.4326} {\bibfield  {journal} {\bibinfo
   {journal} {Phys. Rev. Lett.}\ }\textbf {\bibinfo {volume} {75}},\ \bibinfo
  {pages} {4326} (\bibinfo {year} {1995})}\BibitemShut {NoStop}%
\bibitem [{\citenamefont {Toner}\ \emph {et~al.}(2005)\citenamefont {Toner},
  \citenamefont {Tu},\ and\ \citenamefont {Ramaswamy}}]{Toner05}%
  \BibitemOpen
  \bibfield  {author} {\bibinfo {author} {\bibfnamefont {J.}~\bibnamefont
  {Toner}}, \bibinfo {author} {\bibfnamefont {Y.}~\bibnamefont {Tu}}, \ and\
  \bibinfo {author} {\bibfnamefont {S.}~\bibnamefont {Ramaswamy}},\ }\href@noop
  {} {\bibfield  {journal} {\bibinfo  {journal} {Ann. Phys.}\ }\textbf
  {\bibinfo {volume} {318}},\ \bibinfo {pages} {170} (\bibinfo {year}
  {2005})}\BibitemShut {NoStop}%
\bibitem [{\citenamefont {Fruchart}\ \emph {et~al.}(2021)\citenamefont
  {Fruchart}, \citenamefont {Hanai}, \citenamefont {Littlewood},\ and\
  \citenamefont {Vitelli}}]{Fruchart21}%
  \BibitemOpen
  \bibfield  {author} {\bibinfo {author} {\bibfnamefont {M.}~\bibnamefont
  {Fruchart}}, \bibinfo {author} {\bibfnamefont {R.}~\bibnamefont {Hanai}},
  \bibinfo {author} {\bibfnamefont {P.~B.}\ \bibnamefont {Littlewood}}, \ and\
  \bibinfo {author} {\bibfnamefont {V.}~\bibnamefont {Vitelli}},\ }\href@noop
  {} {\bibfield  {journal} {\bibinfo  {journal} {Nature}\ }\textbf {\bibinfo
  {volume} {592}},\ \bibinfo {pages} {363} (\bibinfo {year}
  {2021})}\BibitemShut {NoStop}%
\bibitem [{\citenamefont {Diessel}\ \emph {et~al.}(2022)\citenamefont
  {Diessel}, \citenamefont {Diehl},\ and\ \citenamefont
  {Chiocchetta}}]{Diessel22}%
  \BibitemOpen
  \bibfield  {author} {\bibinfo {author} {\bibfnamefont {O.~K.}\ \bibnamefont
  {Diessel}}, \bibinfo {author} {\bibfnamefont {S.}~\bibnamefont {Diehl}}, \
  and\ \bibinfo {author} {\bibfnamefont {A.}~\bibnamefont {Chiocchetta}},\
  }\href {\doibase 10.1103/PhysRevLett.128.070401} {\bibfield  {journal}
  {\bibinfo  {journal} {Phys. Rev. Lett.}\ }\textbf {\bibinfo {volume} {128}},\
  \bibinfo {pages} {070401} (\bibinfo {year} {2022})}\BibitemShut {NoStop}%
\bibitem [{\citenamefont {Bukov}\ \emph {et~al.}(2015)\citenamefont {Bukov},
  \citenamefont {Gopalakrishnan}, \citenamefont {Knap},\ and\ \citenamefont
  {Demler}}]{bukov2015}%
  \BibitemOpen
  \bibfield  {author} {\bibinfo {author} {\bibfnamefont {M.}~\bibnamefont
  {Bukov}}, \bibinfo {author} {\bibfnamefont {S.}~\bibnamefont
  {Gopalakrishnan}}, \bibinfo {author} {\bibfnamefont {M.}~\bibnamefont
  {Knap}}, \ and\ \bibinfo {author} {\bibfnamefont {E.}~\bibnamefont
  {Demler}},\ }\href {\doibase 10.1103/PhysRevLett.115.205301} {\bibfield
  {journal} {\bibinfo  {journal} {Phys. Rev. Lett.}\ }\textbf {\bibinfo
  {volume} {115}},\ \bibinfo {pages} {205301} (\bibinfo {year}
  {2015})}\BibitemShut {NoStop}%
\bibitem [{\citenamefont {Landau}\ and\ \citenamefont
  {Lifshitz}(2013)}]{landau2013statistical}%
  \BibitemOpen
  \bibfield  {author} {\bibinfo {author} {\bibfnamefont {L.~D.}\ \bibnamefont
  {Landau}}\ and\ \bibinfo {author} {\bibfnamefont {E.~M.}\ \bibnamefont
  {Lifshitz}},\ }\href@noop {} {\emph {\bibinfo {title} {Statistical Physics:
  Volume 5}}},\ Vol.~\bibinfo {volume} {5}\ (\bibinfo  {publisher} {Elsevier},\
  \bibinfo {year} {2013})\BibitemShut {NoStop}%
\bibitem [{\citenamefont {Stokes}\ \emph {et~al.}(1993)\citenamefont {Stokes},
  \citenamefont {Hatch},\ and\ \citenamefont {Nelson}}]{Stokes1993}%
  \BibitemOpen
  \bibfield  {author} {\bibinfo {author} {\bibfnamefont {H.~T.}\ \bibnamefont
  {Stokes}}, \bibinfo {author} {\bibfnamefont {D.~M.}\ \bibnamefont {Hatch}}, \
  and\ \bibinfo {author} {\bibfnamefont {H.~M.}\ \bibnamefont {Nelson}},\
  }\href {\doibase 10.1103/PhysRevB.47.9080} {\bibfield  {journal} {\bibinfo
  {journal} {Phys. Rev. B}\ }\textbf {\bibinfo {volume} {47}},\ \bibinfo
  {pages} {9080} (\bibinfo {year} {1993})}\BibitemShut {NoStop}%
\bibitem [{\citenamefont {Kopsky}\ and\ \citenamefont
  {Sannikov}(1977)}]{Kopsky_1977}%
  \BibitemOpen
  \bibfield  {author} {\bibinfo {author} {\bibfnamefont {V.}~\bibnamefont
  {Kopsky}}\ and\ \bibinfo {author} {\bibfnamefont {D.~G.}\ \bibnamefont
  {Sannikov}},\ }\href {\doibase 10.1088/0022-3719/10/21/021} {\bibfield
  {journal} {\bibinfo  {journal} {Journal of Physics C: Solid State Physics}\
  }\textbf {\bibinfo {volume} {10}},\ \bibinfo {pages} {4347} (\bibinfo {year}
  {1977})}\BibitemShut {NoStop}%
\bibitem [{\citenamefont {Moriya}(1960)}]{moriya1960anisotropic}%
  \BibitemOpen
  \bibfield  {author} {\bibinfo {author} {\bibfnamefont {T.}~\bibnamefont
  {Moriya}},\ }\href@noop {} {\bibfield  {journal} {\bibinfo  {journal}
  {Physical review}\ }\textbf {\bibinfo {volume} {120}},\ \bibinfo {pages} {91}
  (\bibinfo {year} {1960})}\BibitemShut {NoStop}%
\bibitem [{\citenamefont
  {Dzyaloshinsky}(1958)}]{dzyaloshinsky1958thermodynamic}%
  \BibitemOpen
  \bibfield  {author} {\bibinfo {author} {\bibfnamefont {I.}~\bibnamefont
  {Dzyaloshinsky}},\ }\href@noop {} {\bibfield  {journal} {\bibinfo  {journal}
  {Journal of physics and chemistry of solids}\ }\textbf {\bibinfo {volume}
  {4}},\ \bibinfo {pages} {241} (\bibinfo {year} {1958})}\BibitemShut {NoStop}%
\bibitem [{\citenamefont {Mineev}\ and\ \citenamefont
  {Samokhin}(2008)}]{mineev2008}%
  \BibitemOpen
  \bibfield  {author} {\bibinfo {author} {\bibfnamefont {V.~P.}\ \bibnamefont
  {Mineev}}\ and\ \bibinfo {author} {\bibfnamefont {K.~V.}\ \bibnamefont
  {Samokhin}},\ }\href {\doibase 10.1103/PhysRevB.78.144503} {\bibfield
  {journal} {\bibinfo  {journal} {Phys. Rev. B}\ }\textbf {\bibinfo {volume}
  {78}},\ \bibinfo {pages} {144503} (\bibinfo {year} {2008})}\BibitemShut
  {NoStop}%
\bibitem [{\citenamefont {Nagashima}\ \emph {et~al.}(2024)\citenamefont
  {Nagashima}, \citenamefont {Tian}, \citenamefont {Haenel}, \citenamefont
  {Tsuji},\ and\ \citenamefont {Manske}}]{manske2024}%
  \BibitemOpen
  \bibfield  {author} {\bibinfo {author} {\bibfnamefont {R.}~\bibnamefont
  {Nagashima}}, \bibinfo {author} {\bibfnamefont {S.}~\bibnamefont {Tian}},
  \bibinfo {author} {\bibfnamefont {R.}~\bibnamefont {Haenel}}, \bibinfo
  {author} {\bibfnamefont {N.}~\bibnamefont {Tsuji}}, \ and\ \bibinfo {author}
  {\bibfnamefont {D.}~\bibnamefont {Manske}},\ }\href {\doibase
  10.1103/PhysRevResearch.6.013120} {\bibfield  {journal} {\bibinfo  {journal}
  {Phys. Rev. Res.}\ }\textbf {\bibinfo {volume} {6}},\ \bibinfo {pages}
  {013120} (\bibinfo {year} {2024})}\BibitemShut {NoStop}%
\bibitem [{\citenamefont {Son}\ \emph {et~al.}(2019)\citenamefont {Son},
  \citenamefont {Park}, \citenamefont {Kim}, \citenamefont {Cho}, \citenamefont
  {Kim}, \citenamefont {Sandilands}, \citenamefont {Sohn}, \citenamefont
  {Park}, \citenamefont {Moon},\ and\ \citenamefont
  {Noh}}]{son2019unconventional}%
  \BibitemOpen
  \bibfield  {author} {\bibinfo {author} {\bibfnamefont {J.}~\bibnamefont
  {Son}}, \bibinfo {author} {\bibfnamefont {B.~C.}\ \bibnamefont {Park}},
  \bibinfo {author} {\bibfnamefont {C.~H.}\ \bibnamefont {Kim}}, \bibinfo
  {author} {\bibfnamefont {H.}~\bibnamefont {Cho}}, \bibinfo {author}
  {\bibfnamefont {S.~Y.}\ \bibnamefont {Kim}}, \bibinfo {author} {\bibfnamefont
  {L.~J.}\ \bibnamefont {Sandilands}}, \bibinfo {author} {\bibfnamefont
  {C.}~\bibnamefont {Sohn}}, \bibinfo {author} {\bibfnamefont {J.-G.}\
  \bibnamefont {Park}}, \bibinfo {author} {\bibfnamefont {S.~J.}\ \bibnamefont
  {Moon}}, \ and\ \bibinfo {author} {\bibfnamefont {T.~W.}\ \bibnamefont
  {Noh}},\ }\href@noop {} {\bibfield  {journal} {\bibinfo  {journal} {npj
  Quantum materials}\ }\textbf {\bibinfo {volume} {4}},\ \bibinfo {pages} {17}
  (\bibinfo {year} {2019})}\BibitemShut {NoStop}%
\bibitem [{\citenamefont {Barman}\ \emph {et~al.}(2021)\citenamefont {Barman},
  \citenamefont {Gubbiotti}, \citenamefont {Ladak}, \citenamefont {Adeyeye},
  \citenamefont {Krawczyk}, \citenamefont {Gräfe}, \citenamefont {Adelmann},
  \citenamefont {Cotofana}, \citenamefont {Naeemi}, \citenamefont {Vasyuchka},
  \citenamefont {Hillebrands}, \citenamefont {Nikitov}, \citenamefont {Yu},
  \citenamefont {Grundler}, \citenamefont {Sadovnikov}, \citenamefont
  {Grachev}, \citenamefont {Sheshukova}, \citenamefont {Duquesne},
  \citenamefont {Marangolo}, \citenamefont {Csaba}, \citenamefont {Porod},
  \citenamefont {Demidov}, \citenamefont {Urazhdin}, \citenamefont
  {Demokritov}, \citenamefont {Albisetti}, \citenamefont {Petti}, \citenamefont
  {Bertacco}, \citenamefont {Schultheiss}, \citenamefont {Kruglyak},
  \citenamefont {Poimanov}, \citenamefont {Sahoo}, \citenamefont {Sinha},
  \citenamefont {Yang}, \citenamefont {Münzenberg}, \citenamefont {Moriyama},
  \citenamefont {Mizukami}, \citenamefont {Landeros}, \citenamefont {Gallardo},
  \citenamefont {Carlotti}, \citenamefont {Kim}, \citenamefont {Stamps},
  \citenamefont {Camley}, \citenamefont {Rana}, \citenamefont {Otani},
  \citenamefont {Yu}, \citenamefont {Yu}, \citenamefont {Bauer}, \citenamefont
  {Back}, \citenamefont {Uhrig}, \citenamefont {Dobrovolskiy}, \citenamefont
  {Budinska}, \citenamefont {Qin}, \citenamefont {van Dijken}, \citenamefont
  {Chumak}, \citenamefont {Khitun}, \citenamefont {Nikonov}, \citenamefont
  {Young}, \citenamefont {Zingsem},\ and\ \citenamefont
  {Winklhofer}}]{Barman_2021}%
  \BibitemOpen
  \bibfield  {author} {\bibinfo {author} {\bibfnamefont {A.}~\bibnamefont
  {Barman}}, \bibinfo {author} {\bibfnamefont {G.}~\bibnamefont {Gubbiotti}},
  \bibinfo {author} {\bibfnamefont {S.}~\bibnamefont {Ladak}}, \bibinfo
  {author} {\bibfnamefont {A.~O.}\ \bibnamefont {Adeyeye}}, \bibinfo {author}
  {\bibfnamefont {M.}~\bibnamefont {Krawczyk}}, \bibinfo {author}
  {\bibfnamefont {J.}~\bibnamefont {Gräfe}}, \bibinfo {author} {\bibfnamefont
  {C.}~\bibnamefont {Adelmann}}, \bibinfo {author} {\bibfnamefont
  {S.}~\bibnamefont {Cotofana}}, \bibinfo {author} {\bibfnamefont
  {A.}~\bibnamefont {Naeemi}}, \bibinfo {author} {\bibfnamefont {V.~I.}\
  \bibnamefont {Vasyuchka}}, \bibinfo {author} {\bibfnamefont {B.}~\bibnamefont
  {Hillebrands}}, \bibinfo {author} {\bibfnamefont {S.~A.}\ \bibnamefont
  {Nikitov}}, \bibinfo {author} {\bibfnamefont {H.}~\bibnamefont {Yu}},
  \bibinfo {author} {\bibfnamefont {D.}~\bibnamefont {Grundler}}, \bibinfo
  {author} {\bibfnamefont {A.~V.}\ \bibnamefont {Sadovnikov}}, \bibinfo
  {author} {\bibfnamefont {A.~A.}\ \bibnamefont {Grachev}}, \bibinfo {author}
  {\bibfnamefont {S.~E.}\ \bibnamefont {Sheshukova}}, \bibinfo {author}
  {\bibfnamefont {J.-Y.}\ \bibnamefont {Duquesne}}, \bibinfo {author}
  {\bibfnamefont {M.}~\bibnamefont {Marangolo}}, \bibinfo {author}
  {\bibfnamefont {G.}~\bibnamefont {Csaba}}, \bibinfo {author} {\bibfnamefont
  {W.}~\bibnamefont {Porod}}, \bibinfo {author} {\bibfnamefont {V.~E.}\
  \bibnamefont {Demidov}}, \bibinfo {author} {\bibfnamefont {S.}~\bibnamefont
  {Urazhdin}}, \bibinfo {author} {\bibfnamefont {S.~O.}\ \bibnamefont
  {Demokritov}}, \bibinfo {author} {\bibfnamefont {E.}~\bibnamefont
  {Albisetti}}, \bibinfo {author} {\bibfnamefont {D.}~\bibnamefont {Petti}},
  \bibinfo {author} {\bibfnamefont {R.}~\bibnamefont {Bertacco}}, \bibinfo
  {author} {\bibfnamefont {H.}~\bibnamefont {Schultheiss}}, \bibinfo {author}
  {\bibfnamefont {V.~V.}\ \bibnamefont {Kruglyak}}, \bibinfo {author}
  {\bibfnamefont {V.~D.}\ \bibnamefont {Poimanov}}, \bibinfo {author}
  {\bibfnamefont {S.}~\bibnamefont {Sahoo}}, \bibinfo {author} {\bibfnamefont
  {J.}~\bibnamefont {Sinha}}, \bibinfo {author} {\bibfnamefont
  {H.}~\bibnamefont {Yang}}, \bibinfo {author} {\bibfnamefont {M.}~\bibnamefont
  {Münzenberg}}, \bibinfo {author} {\bibfnamefont {T.}~\bibnamefont
  {Moriyama}}, \bibinfo {author} {\bibfnamefont {S.}~\bibnamefont {Mizukami}},
  \bibinfo {author} {\bibfnamefont {P.}~\bibnamefont {Landeros}}, \bibinfo
  {author} {\bibfnamefont {R.~A.}\ \bibnamefont {Gallardo}}, \bibinfo {author}
  {\bibfnamefont {G.}~\bibnamefont {Carlotti}}, \bibinfo {author}
  {\bibfnamefont {J.-V.}\ \bibnamefont {Kim}}, \bibinfo {author} {\bibfnamefont
  {R.~L.}\ \bibnamefont {Stamps}}, \bibinfo {author} {\bibfnamefont {R.~E.}\
  \bibnamefont {Camley}}, \bibinfo {author} {\bibfnamefont {B.}~\bibnamefont
  {Rana}}, \bibinfo {author} {\bibfnamefont {Y.}~\bibnamefont {Otani}},
  \bibinfo {author} {\bibfnamefont {W.}~\bibnamefont {Yu}}, \bibinfo {author}
  {\bibfnamefont {T.}~\bibnamefont {Yu}}, \bibinfo {author} {\bibfnamefont
  {G.~E.~W.}\ \bibnamefont {Bauer}}, \bibinfo {author} {\bibfnamefont
  {C.}~\bibnamefont {Back}}, \bibinfo {author} {\bibfnamefont {G.~S.}\
  \bibnamefont {Uhrig}}, \bibinfo {author} {\bibfnamefont {O.~V.}\ \bibnamefont
  {Dobrovolskiy}}, \bibinfo {author} {\bibfnamefont {B.}~\bibnamefont
  {Budinska}}, \bibinfo {author} {\bibfnamefont {H.}~\bibnamefont {Qin}},
  \bibinfo {author} {\bibfnamefont {S.}~\bibnamefont {van Dijken}}, \bibinfo
  {author} {\bibfnamefont {A.~V.}\ \bibnamefont {Chumak}}, \bibinfo {author}
  {\bibfnamefont {A.}~\bibnamefont {Khitun}}, \bibinfo {author} {\bibfnamefont
  {D.~E.}\ \bibnamefont {Nikonov}}, \bibinfo {author} {\bibfnamefont {I.~A.}\
  \bibnamefont {Young}}, \bibinfo {author} {\bibfnamefont {B.~W.}\ \bibnamefont
  {Zingsem}}, \ and\ \bibinfo {author} {\bibfnamefont {M.}~\bibnamefont
  {Winklhofer}},\ }\href {\doibase 10.1088/1361-648X/abec1a} {\bibfield
  {journal} {\bibinfo  {journal} {Journal of Physics: Condensed Matter}\
  }\textbf {\bibinfo {volume} {33}},\ \bibinfo {pages} {413001} (\bibinfo
  {year} {2021})}\BibitemShut {NoStop}%
\bibitem [{\citenamefont {Floris}\ \emph {et~al.}(2011)\citenamefont {Floris},
  \citenamefont {de~Gironcoli}, \citenamefont {Gross},\ and\ \citenamefont
  {Cococcioni}}]{niophonons}%
  \BibitemOpen
  \bibfield  {author} {\bibinfo {author} {\bibfnamefont {A.}~\bibnamefont
  {Floris}}, \bibinfo {author} {\bibfnamefont {S.}~\bibnamefont
  {de~Gironcoli}}, \bibinfo {author} {\bibfnamefont {E.~K.~U.}\ \bibnamefont
  {Gross}}, \ and\ \bibinfo {author} {\bibfnamefont {M.}~\bibnamefont
  {Cococcioni}},\ }\href {\doibase 10.1103/PhysRevB.84.161102} {\bibfield
  {journal} {\bibinfo  {journal} {Phys. Rev. B}\ }\textbf {\bibinfo {volume}
  {84}},\ \bibinfo {pages} {161102} (\bibinfo {year} {2011})}\BibitemShut
  {NoStop}%
\bibitem [{\citenamefont {Hutchings}\ and\ \citenamefont
  {Samuelsen}(1972)}]{niomagnons}%
  \BibitemOpen
  \bibfield  {author} {\bibinfo {author} {\bibfnamefont {M.~T.}\ \bibnamefont
  {Hutchings}}\ and\ \bibinfo {author} {\bibfnamefont {E.~J.}\ \bibnamefont
  {Samuelsen}},\ }\href {\doibase 10.1103/PhysRevB.6.3447} {\bibfield
  {journal} {\bibinfo  {journal} {Phys. Rev. B}\ }\textbf {\bibinfo {volume}
  {6}},\ \bibinfo {pages} {3447} (\bibinfo {year} {1972})}\BibitemShut
  {NoStop}%
\bibitem [{\citenamefont {Costa}\ \emph {et~al.}(2020)\citenamefont {Costa},
  \citenamefont {Santos}, \citenamefont {Peres},\ and\ \citenamefont
  {Fern{\'a}ndez-Rossier}}]{costa2020topological}%
  \BibitemOpen
  \bibfield  {author} {\bibinfo {author} {\bibfnamefont {A.~T.}\ \bibnamefont
  {Costa}}, \bibinfo {author} {\bibfnamefont {D.~L.~R.}\ \bibnamefont
  {Santos}}, \bibinfo {author} {\bibfnamefont {N.~M.}\ \bibnamefont {Peres}}, \
  and\ \bibinfo {author} {\bibfnamefont {J.}~\bibnamefont
  {Fern{\'a}ndez-Rossier}},\ }\href@noop {} {\bibfield  {journal} {\bibinfo
  {journal} {2D Materials}\ }\textbf {\bibinfo {volume} {7}},\ \bibinfo {pages}
  {045031} (\bibinfo {year} {2020})}\BibitemShut {NoStop}%
\bibitem [{\citenamefont {Kleiner}\ \emph {et~al.}(2021)\citenamefont
  {Kleiner}, \citenamefont {Zhou}, \citenamefont {Dorsch}, \citenamefont
  {Zhang}, \citenamefont {Koelle},\ and\ \citenamefont {Jin}}]{Kleiner21}%
  \BibitemOpen
  \bibfield  {author} {\bibinfo {author} {\bibfnamefont {R.}~\bibnamefont
  {Kleiner}}, \bibinfo {author} {\bibfnamefont {Z.}~\bibnamefont {Zhou}},
  \bibinfo {author} {\bibfnamefont {E.}~\bibnamefont {Dorsch}}, \bibinfo
  {author} {\bibfnamefont {X.}~\bibnamefont {Zhang}}, \bibinfo {author}
  {\bibfnamefont {D.}~\bibnamefont {Koelle}}, \ and\ \bibinfo {author}
  {\bibfnamefont {D.}~\bibnamefont {Jin}},\ }\href@noop {} {\bibfield
  {journal} {\bibinfo  {journal} {Nature Communications}\ }\textbf {\bibinfo
  {volume} {12}},\ \bibinfo {pages} {6038} (\bibinfo {year}
  {2021})}\BibitemShut {NoStop}%
\bibitem [{\citenamefont {Burns}\ and\ \citenamefont
  {Scott}(1973)}]{burns1973lattice}%
  \BibitemOpen
  \bibfield  {author} {\bibinfo {author} {\bibfnamefont {G.}~\bibnamefont
  {Burns}}\ and\ \bibinfo {author} {\bibfnamefont {B.~A.}\ \bibnamefont
  {Scott}},\ }\href@noop {} {\bibfield  {journal} {\bibinfo  {journal}
  {Physical Review B}\ }\textbf {\bibinfo {volume} {7}},\ \bibinfo {pages}
  {3088} (\bibinfo {year} {1973})}\BibitemShut {NoStop}%
\bibitem [{\citenamefont {Tchernyshyov}(2024)}]{tchernyshyov2024field}%
  \BibitemOpen
  \bibfield  {author} {\bibinfo {author} {\bibfnamefont {O.}~\bibnamefont
  {Tchernyshyov}},\ }\href@noop {} {\enquote {\bibinfo {title} {Field theory of
  collinear and noncollinear magnetic order},}\ } (\bibinfo {year} {2024}),\
  \Eprint {http://arxiv.org/abs/2401.07171} {arXiv:2401.07171 [cond-mat.other]}
  \BibitemShut {NoStop}%
\bibitem [{\citenamefont {Dormand}\ and\ \citenamefont
  {Prince}(1980)}]{dormand1980family}%
  \BibitemOpen
  \bibfield  {author} {\bibinfo {author} {\bibfnamefont {J.~R.}\ \bibnamefont
  {Dormand}}\ and\ \bibinfo {author} {\bibfnamefont {P.~J.}\ \bibnamefont
  {Prince}},\ }\href@noop {} {\bibfield  {journal} {\bibinfo  {journal}
  {Journal of computational and applied mathematics}\ }\textbf {\bibinfo
  {volume} {6}},\ \bibinfo {pages} {19} (\bibinfo {year} {1980})}\BibitemShut
  {NoStop}%
\bibitem [{\citenamefont {Basini}\ \emph {et~al.}(2024)\citenamefont {Basini},
  \citenamefont {Pancaldi}, \citenamefont {Wehinger}, \citenamefont {Udina},
  \citenamefont {Unikandanunni}, \citenamefont {Tadano}, \citenamefont
  {Hoffmann}, \citenamefont {Balatsky},\ and\ \citenamefont
  {Bonetti}}]{Basini2024}%
  \BibitemOpen
  \bibfield  {author} {\bibinfo {author} {\bibfnamefont {M.}~\bibnamefont
  {Basini}}, \bibinfo {author} {\bibfnamefont {M.}~\bibnamefont {Pancaldi}},
  \bibinfo {author} {\bibfnamefont {B.}~\bibnamefont {Wehinger}}, \bibinfo
  {author} {\bibfnamefont {M.}~\bibnamefont {Udina}}, \bibinfo {author}
  {\bibfnamefont {V.}~\bibnamefont {Unikandanunni}}, \bibinfo {author}
  {\bibfnamefont {T.}~\bibnamefont {Tadano}}, \bibinfo {author} {\bibfnamefont
  {M.}~\bibnamefont {Hoffmann}}, \bibinfo {author} {\bibfnamefont
  {A.}~\bibnamefont {Balatsky}}, \ and\ \bibinfo {author} {\bibfnamefont
  {S.}~\bibnamefont {Bonetti}},\ }\href@noop {} {\bibfield  {journal} {\bibinfo
   {journal} {Nature}\ ,\ \bibinfo {pages} {1}} (\bibinfo {year}
  {2024})}\BibitemShut {NoStop}%
\end{thebibliography}%
\clearpage
\onecolumngrid
\begin{center}
\large \textbf{Supplementary Information}
\end{center}
\renewcommand\thesection{S\arabic{section}}
\renewcommand\theequation{S\arabic{equation}}
\renewcommand\theequation{S\arabic{equation}}
\setcounter{equation}{0}
\section{Effective potential}
In the main text, we presented Eq.~1 followed by some approximations. We begin by clarifying that the terms presented in Eq.~1 are the leading order terms in the interaction between $Q$, $P$. The bilinear coupling $Q_i P_i$ is not allowed when $P$ and $Q$ have different symmetry, such as the cases depicted in Fig. 1 (a,b). Even if $P$ and $Q$ do have the same symmetry, its effect is only to renormalize the effective charge of the $Q$, $P$ modes \cite{Zhuang23}. We set the charge of $Q$ mode to $1$. The next-to-leading order contributions are precisely those outlined in Eq.~1. The approximation taken in the main text involves $\xi' = 0$. This is justified since the $Q$ mode is driven resonantly at frequency $\Omega$. As a result, $Q^2$ (the term in $\xi'$) oscillates at either $0,2\Omega$. Neither will lead to a parametric excitation of $P$, and specifically since the dispersion of $P$ is assumed to be below the $Q$ mode. 

For the gradient term $\delta_{ijkl}$ we can easily show that it matches the form of the coupling we presented in the main text. Since $Q$ is spatially uniform and the interaction is overall real, the gradient term may be written as $\delta_{ijkl}Q_i \partial_k (P_j P_l)$. The Fourier transform of this at momentum $q$ is $\delta_{ijkl} q_k Q_i P_{j,q}P_{l,-q}$. This shows the general form of the coupling is, after absorbing all additional  of indices into the coupling function $g(q)$, $g(q) Q_i P_j P_k$.

\section{Dimensions in the equation of motion and damping}

\resp{For clarity, we reintroduce the coupled equations of the main text with their dimension-full analogues for the proper estimation of their size.}

\begin{align}
    M_Q \ddot{Q} + \beta \dot{Q} + M_Q \Omega^2 Q = Z e -2 \gamma P^2, \\
     M_P \ddot{P} + \beta \dot{P} + M_P \Omega_P^2 P = -2 \gamma Q P,
\end{align}
\resp{
For simplicity we dropped the self-interaction nonlinearties for now, but the process of restoring dimensions to their couplings in entirely identical. 
In this form, $Q$ carries units of displacement (length), and $M_{Q/P}$ is the mass. $\beta$ is then a damping term with dimensions mass $\times$ frequency. The dimensionless $\beta_{dimensionless}$ value we use in the main text corresponds to $\beta / M \Omega$ in the equation above. 
This parameter can be extracted experimentally from the spectroscopic linewidth of the phonon $\Gamma = \frac{\beta}{2M}$ and its frequency $\Omega$. 
The ratio of the phonon linewidth in PbTiO\textsubscript{3} \cite{burns1973lattice} to the energy of the $E_1$ (TO) mode that we assume is coupled to the electric field (the $Q$ mode in the model) is $\Gamma/\Omega = 0.15$, given that $\Omega \sim 74 \textrm{cm}^{-1}$ and $\Gamma \sim 12 \textrm{cm}^{-1}$ (see Fig. 11 in Ref.~\cite{burns1973lattice}, at roughly $T = 400 C$). Overall we find good agreement with our chosen value of $\beta$, given the ratio of $\beta/M\Omega = 0.1$ chosen in the main text (Fig. 2).
}

\section{Details of analytical calculations for the order onset}

In general, when both off-resonant driving and deviation from $q^2_0 = ((\Omega/2)^2-\omega_P^2)/v_P^2$ condition are allowed one gets the following equations:
\begin{align}
\begin{gathered}
    (\delta \omega_Q^2- i \Omega \beta) Q_{0, \Omega} = -E_0/2 + 2\gamma P_{q_0,\Omega/2} P^*_{q_0,-\Omega/2}, \\
(\delta \omega_P^2 - i\beta \Omega/2) P_{q_0, \Omega/2} = 2 \gamma Q_{0,\Omega} P_{q_0,-\Omega/2},\\
        (\delta \omega_Q^2+ i \Omega \beta) Q_{0, \Omega}^* = -E_0/2 + 2\gamma P_{q_0,-\Omega/2} P^*_{q_0,\Omega/2}, \\
    (\delta \omega_P^2 + i\beta \Omega/2) P_{q_0, -\Omega/2} = 2 \gamma Q_{0,\Omega}^* P_{q_0,\Omega/2}.
\end{gathered}
\end{align}
Comparing second and third equations implies that $|P_{q_0,\Omega/2}|=|P_{q_0,-\Omega/2}|$ holds in the general case too. Solving for $Q_{0,\Omega}$ in second equation and introducing the result into the first one, we get the following equation:
\begin{align}
    e^{i(\varphi_{\Omega} - \varphi_{-\Omega})} \left[-\frac{(\delta \omega_Q^2 -i \Omega \beta)(\delta\omega_P^2 - i\beta \Omega/2)}{2\gamma} +2 \gamma |P_{q_0,\Omega/2}|^2\right] = E_0/2,
    \label{sup:eq:self-cons}
\end{align}
which coincides with the one in main text for $\delta\omega_P,\delta \omega_Q =0$. As the r.h.s. of \eqref{sup:eq:self-cons} is purely real, the phase difference $\varphi_{\Omega} - \varphi_{-\Omega}$ on the l.h.s. has to be chosen to cancel that of the term in the bracket. Then, the result reduced to:
\begin{equation}
\sqrt{\frac{(2\delta\omega_P^2+\delta\omega_Q^2)^2\Omega^2\beta^2}{16\gamma^2} + (2\gamma |P_{q_0,\Omega/2}|^2+ \beta^2\Omega^2/4\gamma-\delta\omega_Q^2\delta\omega_P^2/2\gamma)^2} = E_0/2,
    \label{sup:eq:self-cons_real}
\end{equation}
resulting in the threshold field being:
\begin{equation}
\begin{gathered}
    E_c = 2\sqrt{\frac{(2\delta\omega_P^2+\delta\omega_Q^2)^2\Omega^2\beta^2}{16\gamma^2} + (\beta^2\Omega^2/4\gamma-\delta\omega_Q^2\delta\omega_P^2/2\gamma)^2}=
    \\
    =
    2\sqrt{\frac{(\beta \Omega)^4}{16\gamma^2}
    +\frac{\delta\omega_P^4 \Omega^2\beta^2}{4\gamma^2}
    +\frac{\delta\omega_Q^4 \Omega^2\beta^2}{16\gamma^2}
    +
    \frac{\delta\omega_P^4  \delta\omega_Q^4}{4\gamma^2}
    }.
    \end{gathered}
\end{equation}

Thus, the threshold field becomes larger for both detuned drive $\delta\omega_Q$ and and modes away from parametric condition $\delta \omega_P$. Due to non-linearity of the system one expects then only modes with $\delta \omega_P = 0$ to be realized for a given $\delta \omega_Q$.

\section{Phenomenological models for the mode coupling}

\subsection{Ferroelectrics}

Consider a tetragonal system undergoing a transition to a polar phase; the uniform part of energy density reads

\begin{equation}
\frac{E}{V} = \frac{a}{2} \vec{X}^2+ \frac{b}{4} (\vec{X}^2)^2 +c X_1^2X_2^2,
\end{equation}
where $a<0, b>0$ in the ordered phase and we assume (without the loss of generality achieved by an appropriate rotation of coordinates) that $c>0$. In the ordered phase, one has then either $X_1^2 = |a|/b,X_2=0$ or vice versa. We denote the component that develops the expectation value as $Q$ and the other one as P. Then, one can derive the effective energy density for small deviations of these modes from their equilibrium positions:
\begin{equation}
\frac{\delta E}{V} = |a| \delta Q^2 + \sqrt{|a| b}  \delta Q^3
+
\frac{c |a|}{b} \delta P^2
+
\sqrt{|a| b} \delta Q \delta P^2
+O(\delta X^4).
\end{equation}
Note that the stiffness of the $P$ mode is proportional to $c$; indeed when $c=0$ the system has continuous rotational symmetry and $P$ is the Goldstone mode. From the above, one can also estimate the $\Delta \omega_P^2/\omega_P^2 = \frac{\sqrt{|a| b} \delta Q}{\frac{c |a|}{b}} = \frac{b}{c} \frac{\Delta Q}{Q_0}$, where $Q_0 = \sqrt{|a|/b}$ is the equilibrium displacement of $Q$. Typically, ferroelectrics have a large anisotropy, so $b/c\sim 1$. At the same time, for our mechanism to be operative we need $\omega_Q> 2\omega_P$, which translates into $|a|>4\frac{c |a|}{b}$ and thus $\frac{c}{b}<1/4$ for this model.

\subsection{Magnets}
Our starting point is a Landau-Lifshitz form of dynamics, with damping neglected. 
We consider a two-sublattice antiferromagnet with a DM interaction \cite{moriya1960anisotropic,dzyaloshinsky1958thermodynamic} modulated by phonon displacement. On a lattice, the DM interaction takes the form $D_{ij} (\mathbf{S}_i \times \mathbf{S}_j)$. Allowing $D_{ij}$ to vary in the presence of the optical mode $Q$, and promoting the interaction to the continuum, we have $Q \mathbf{S} \times \frac{\partial \mathbf{S}}{\partial x}$.
A simple illustration of its effect would be the equations of motion for an isotropic 1D antiferromagnet \cite{tchernyshyov2024field}: $\ddot m_{1(2)} -c^2 \partial_{x}^2 m_{1(2)}+(-)\gamma Q \partial_x m_{2(1)}=0$, where $m_{1(2)}$ are the sublattice magnetization components orthogonal to the equilibrium one. The equations reduce to two oscillators with frequencies $\sqrt{c^2q^2\pm \gamma Q q}$, such that $Q$ shifts linearly the magnon energies, allowing for the spatiotemporal order via phonon driving. This idea may open an alternative path for localized magnon generation (due to large $q_0$) in the THz regime, allowing for faster nanoscale magnonic computation \cite{Barman_2021}. Previous work has shown that in NiO, TO phonons \cite{niophonons} are found at about $30-40$ meV, while the magnon band extends from few to $100$ meV \cite{niomagnons}. Another potential application  of our proposal is in accessing the magnon dispersion of 2D materials, such as CrI\textsubscript{3}, where magnons close to the $K$ point  are predicted to posses a topological gap.
\section{Details of material estimates}
Here we present an estimate for the critical fluence $E_c$ for PbTiO\textsubscript{3}; here we propose directly exciting an optical E phonon mode with frequency $\omega_Q \approx 275 \textrm{cm}^{-1}$\cite{Choudhury2008}, that couples to another E optical phonon mode with frequency $\omega_P \approx 90 \textrm{cm}^{-1}$ via parametric resonance. This will induce a spatial order along the $\Gamma-Z$ direction with wavevector $q_0\approx 0.95 \AA^{-1}$. To provide a rough estimate of $E_c$, we assume that $\beta=0.1$, and use the coupling constant between the highest A$_1$ and lowest A$_1$ phonon mode $\gamma\approx -12.4 \textrm{meV}\cdot \textrm{amu}^{-3/2}\cdot \textrm{\AA}^{-3}$ given in Ref. \cite{Subedi2015}; this yields an estimated $E_0\approx 3.55 \textrm{MV}/\textrm{cm}$.
that should be experimentally accessible. 
\section{Details of the numerical simulation}
\resp{
In the numerical framework we adopted, we have solved a system of coupled partial differential equations. The solution proceeds in the following way: The state $P(x,t) \mapsto P(x_i, t_i)$ with the coordinates $x_i$, belonging to a discretized set of size $N = 301$ points. The size of the box for discretization is set to be $L = 1$ for convenience (we have checked that this value does not change the results in any meaningful way). The value of $N$ was chosen in a trade off between numerical accuracy computational tractibility. We then discretize the Laplace operator $\nabla^2 f(x) \approx (f(x_{i+1})-2 f(x_i)+f(x_{i-1}))/( \Delta x ^2)$ where $\Delta x = L/(N) $. Next we apply the fifth order Runge-Kutta algorithm for the time-dependence (left-hand side of the equation of motion) as implemented in Scipy \cite{dormand1980family}. To integrate with the built-in solver (and in order to check for stiffness), we introduce the auxiliary ``velocity" fields which encode the \textit{first} derivatives in time of the fields: $\bar{P}(x,t) = \dot{P}, \bar{Q} = \dot{Q}$. We then decouple the equations into four coupled \textit{first order} differential equations (in time). For example, for the $Q$ mode,}
\begin{align}
\dot{\bar{Q}}_i &= -\beta \bar{Q}_i - \gamma P_i P_i +E(t), \\
\dot{Q}_i &= \bar{Q}_i
\end{align}
\resp{And similarly with $P$. We omitted the contribution of the spatial gradient, for brevity. We initialize the two fields with random initial conditions. The velocity field may also be set to zero at $t = 0$. This does not affect the final result. We have verified the solutions against the possibility of stiffness or uncontrolled behavior.}
\section{Heating and additional mechanisms for breakdown}
\resp{Here, we consider possible mechanisms for breakdown which might affect the onset of order. We consider the role of the breakdown (in connection with previous experiments on $V_2 O_3$): a back of the envelope calculation for the breakdown field  ($E_c$) of PbTiO\textsubscript{3} (as an example) involves the ratio of the gap to the polar axis size, $E_c \sim \Delta/c$, where $c$ is the tetragonal axis in the ferroelectric phase. One finds $E_c \sim 87 \textrm{MV/cm}$ which exceeds by an order of magnitude the estimate that we provide for threshold field needed for our effect in the same material: $\sim 3.55 \textrm{MV/cm}$. We also note that in a recent experiment, THz illumination of SrTiO\textsubscript{3} \cite{Fechner2024}, application of $\sim 60 \textrm{mJ}/\textrm{cm}^2 \approx 2 \textrm{MV/cm}$ the crystal survived without breakdown. This gives us a wider range of applicability and greater flexibility in the intensity of the field. 
As there is a possibility of resistivity switching due to the application of intense electric fields (as was observed e.g. in  $V_2 O_3$) we briefly comment that this effect is typically observed at $\omega \to 0$. We are considering oscillating fields of $\sim 10^{12} \textrm{Hz}$ where linear effects of resistive switching are expected to average out. }

\resp{As for the heating generated by absorption, we refer to a recent estimate \cite{Basini2024} provided for SrTiO\textsubscript{3} (STO) with similar conditions we envisage in an experiment probing our proposed effect \cite{Basini2024}. The temperature increase for a pump with fluence $60 \mu J cm^{-2}$ was estimated in Ref.~\cite{Basini2024} to be $\Delta T =0.04$ K. For the fluence $60 m J cm^{-2}$ which, as demonstrated above, yields sufficient $E$-field amplitude, $\Delta T \approx 40  K$ can be estimated. Compared to the room temperature $T\approx 300$ K such a temperature increase is not expected to dramatically affect the material properties, as both the electronic band gap $>1$ eV and typical longitudinal phonon energies $\sim 100$ meV are much larger than $\Delta T$.}

\end{document}